\documentclass[lettersize,journal]{IEEEtran}
\IEEEoverridecommandlockouts
\usepackage{cite}
\usepackage{amsmath,amssymb,amsfonts}
\usepackage{algorithmic}
\usepackage{graphicx}
\usepackage{textcomp}
\usepackage{xcolor}
\renewcommand{\IEEEQED}{\IEEEQEDopen}
\newtheorem{lemma}{Lemma}

\newtheorem{remark}{Remark}

\begin{document}

\title{Massive MIMO with Dual-Polarized Antennas }

\author{\IEEEauthorblockN{\"Ozgecan \"Ozdogan, Emil Bj\"ornson, \emph{IEEE Fellow}}\\ 
		\thanks{\"O.~\"Ozdogan was with Linköping University, 581 83 Linköping, Sweden. She is now with Ericsson Research, 583 30 Linköping, Sweden (e-mail: ozgecan.ozdogan.senol@ericsson.com).}
		\thanks{E.~Bj\"ornson is with the Department of Computer Science, KTH Royal Institute of Technology, 10044 Stockholm Sweden (Email: emilbjo@kth.se).}
		\thanks{This paper was supported by the Grant 2019-05068 from the Swedish Research Council.}}

\maketitle

\begin{abstract}
This paper considers a single-cell massive MIMO (multiple-input multiple-output) system with dual-polarized antennas at both the base station and users. We study a channel model that includes the key practical aspects that arise when utilizing dual-polarization: channel cross-polar discrimination (XPD) and cross-polar correlations (XPC) at the transmitter and receiver. We derive the achievable uplink and downlink spectral efficiencies (SE) with and without successive interference cancellation (SIC) when using the linear minimum mean squared error (MMSE), zero-forcing (ZF), and maximum ratio (MR) combining/precoding schemes. The expressions depend on the statistical properties of the MMSE channel estimator obtained for the dual-polarized channel model. Closed-form uplink and downlink SE expressions for MR combining/precoding are derived. Using these expressions, we propose power-control algorithms that maximize the uplink and downlink sum SEs under uncorrelated fading but can be used to enhance performance also with correlated fading. We compare the SEs achieved in dual-polarized and uni-polarized setups numerically and evaluate the impact of XPD and XPC conditions. The simulations reveal that dual-polarized setups achieve 40-60\% higher SEs and the gains remain also under severe XPD and XPC. Dual-polarized also systems benefit more from advanced signal processing that compensates for imperfections.
\end{abstract} 


\section{Introduction}

Massive MIMO (multiple-input multiple-output) is the key technology for increasing the spectral efficiency (SE)  in 5G and beyond-5G cellular networks, by virtue of adaptive beamforming and spatial multiplexing \cite{Larsson2014a}. A massive MIMO base station (BS) is equipped with a large number of individually controllable antenna-integrated radios, which can be effectively used to serve tens of user equipments (UEs)  on the same time-frequency resource. Wireless signals are polarized electromagnetic waves and since the electric flux can oscillate in two dimensions perpendicular to the direction that the wave is traveling, there exist two orthogonal polarization dimensions.
Practical BSs and UEs typically utilize  dual-polarized antennas (i.e., two co-located antennas that respond to orthogonal polarizations \cite{Wong2004a}) to squeeze in twice the number of antennas in the same physical enclosure \cite{Asplund20}, and capture components from both polarization dimensions that the arriving signal from the transmitter can have, for diversity and multiplexing purposes.\footnote{It is also possible to use tri-polarized antennas \cite{Andrews2001a}, but that is mainly useful for short-range communications with rich scattering around the transmitter and the receiver, so that signals can propagate in any direction from the transmitter and reach the receiver from any direction.} Nevertheless, the main theory for massive MIMO has been developed for uni-polarized single-antenna users \cite{massivemimobook}.

The channel modeling for dual-polarized channels is substantially more complicated than for conventional uni-polarized channels.
Several measurements and channel models considering dual-polarized antennas are reported in prior literature. In \cite{Shafi2006a, Calcev2007}, the authors provide geometry-based channel models based on measurement campaigns for dual-polarized small-scale  MIMO systems. In addition,  \cite{coldrey2008modeling, oestges2008dual}  provide analytical channel models based on extensive surveys of experimental results for single-user dual-polarized MIMO systems. The mentioned channel models emphasize the need for including different polarization-related properties in different scenarios, thus, there is no model that is well-suited for every scenario. In this paper, each individual channel is modeled along the lines of the widely used model in \cite{coldrey2008modeling} where the two polarizations are statistically symmetric. We select this model since it is analytically tractable, yet it includes the key effects of channel cross-polar discrimination (XPD) and cross-polar receive and transmit correlations (XPC) commonly observed in measurements.

The capacity loss due to the polarization mismatch in a single-user dual-polarized MISO system is analyzed in \cite{joung2014capacity}. The single-user case is also considered in \cite{Qian2018} with focus using an angular channel decomposition for channel estimation.
Massive MIMO scenarios with dual-polarized BS antennas and multiple users, each equipped with a single uni-polarized antenna, are considered in \cite{park2015multi,Khalilsarai2022}. Both papers considers frequency division duplex (FDD) mode.
The energy efficiency of a setup similar to \cite{park2015multi,Khalilsarai2022} is evaluated in \cite{Yin2020}, while polarization leakage between the antennas is ignored (which makes the channels with different polarizations orthogonal). A distributed FDD massive MIMO system, where each user and  distributed antenna port have a single uni-polarized antenna, is considered in \cite{Park2020}. A multi-user massive MIMO system with non-orthogonal multiple access is considered in \cite{Sena2019}. The users are grouped  so that the users that are in the same group share the same spatial correlation matrix, which is a simplifying assumption. Some other recent papers related to dual-polarized antennas are \cite{Hemadeh2020,Zhang2020p,Khalilsarai2020}.  In \cite{Hemadeh2020} and \cite{Zhang2020p},  polarization-based modulation schemes are proposed. Furthermore, reconfigurable dual-polarized antennas that are able to change their polarization states are considered in \cite{Hemadeh2020}.  The authors in \cite{Khalilsarai2020} consider the channel correlation matrix estimation problem.

The canonical form of massive MIMO operates in time-division duplex (TDD) mode and acquires channel state information (CSI) for uplink and downlink transmissions by using uplink pilot signaling and uplink-downlink channel reciprocity \cite{massivemimobook}. 
This paper evaluates a single-cell massive MIMO system with multiple multi-antenna users operating in TDD mode. This 5G scenario is not covered in the aforementioned previous works, which focus on a single dual-polarized user, FDD mode, and/or multiple uni-polarized users.
  The main contributions are:
\begin{itemize}
	\item We study a multi-user massive MIMO scenario with dual-polarized antennas at both the BS and UE sides, and spatial correlation at both sides while following the polarization modeling approach from \cite{coldrey2008modeling}. To the best of our knowledge, this case has previously only been studied in the simplistic case with equal transmit spatial correlation matrices among the users \cite{Sena2019}.
	
		\item  We analyze uplink and downlink achievable SEs with and without successive interference cancellation (SIC) for the linear minimum mean square error (MMSE), zero-forcing (ZF) and maximum ratio (MR) combining/precoding schemes. 
	
	\item  We particularize the classical MMSE channel estimator for the considered dual-polarized system model. Using the resulting covariance matrices, we compute  closed-form uplink and downlink SE expressions when the estimates are used for MR combining/precoding.
	
	\item  Based on the closed-form SE expressions, we  provide power control algorithms to maximize the uplink and downlink sum SEs.

	\item  The dual-polarized and uni-polarized antenna setups are compared numerically. The impact of power control, XPD, and XPC  on uplink and downlink SEs are also evaluated.
\end{itemize}
The conference version of this paper \cite{WSA2021} only considered
the downlink transmission with MMSE-SIC scheme and no power control.

\textit{Reproducible research:} All the simulation results can be
reproduced using the Matlab code and data files available at:
https://github.com/emilbjornson/dual-polarization 

\textit{Notation}: Lower and upper case bold letters are used for vectors and matrices.  The transpose and Hermitian transpose of a matrix $\mathbf{A}$ are written as $\mathbf{A}^T$ and $\mathbf{A}^H$, respectively. The superscript $(.)^*$ denotes the complex conjugate operation. The $M \times M $-dimensional matrix with the diagonal elements $d_1, d_2, \dots, d_M$ is denoted as $\mathrm{diag}\left(d_1, d_2, \dots, d_M \right) $. The diagonal elements of a matrix $\mathbf{D}$ are extracted  to a $M \times 1$ vector as $\mathrm{diag}(\mathbf{D})=[d_1, d_2, \dots, d_M]^T$.  The expectation of a random variable $X$ is denoted by $\mathbb{E}\left\lbrace X\right\rbrace$.

\enlargethispage{3mm}

\section{System model with dual-polarized antennas}

We consider a single-cell massive MIMO system with  $\frac{M}{2}$  dual-polarized  antennas at the BS and $K$ UEs, each equipped with a single dual-polarized antenna.  Each dual-polarized antenna is composed of one vertical (V) and one horizontal (H) polarized antennas that are co-located.\footnote{The analysis holds for any set of two orthogonally polarizations, which could also be slanted $\pm 45^\circ$ linear polarizations or clockwise/counter-clockwise circular polarizations. We just refer to them as V and H polarized for notational convenience, but the analytical results are applicable to all of these cases whenever the same spatial correlation occurs for both polarizations.} A V/H polarized antenna emits and receives electromagnetic waves whose electric field oscillates in the V/H plane. Thus, the BS has $M$ antennas in total and each UE has two antennas, if one counts the number of ports (inputs/outputs, radio-frequency chains).
Note that an array with a given aperture can accommodate twice as many antennas if dual-polarized antennas are utilized, compared to uni-polarized antennas. 
The system operates in TDD mode and we consider the standard block fading model \cite{massivemimobook}, where the channels are static and frequency-flat within a coherence time-frequency block, and varies independently between blocks. We let $\tau_c$ denote the number of transmission samples per block.
{Since the majority of the traffic in cellular networks is generated by indoor users \cite{EricssonMobility}, we will assume a non-line-of-sight (NLOS) propagation model. The analysis can be generalized to line-of-sight scenarios using the methodology in \cite{ozdogan2019massive} but at the expense of more complicated formulas that provide less intuition.}

Extending  \cite{coldrey2008modeling} and \cite{joung2014capacity} to  $\frac{M}{2}$ dual-polarized antennas and multiple UEs, the propagation channel of the UE $k$ is 
\begin{align}\label{eq4} 
\!	\mathbf{H}_k &= \begin{bmatrix}
		\mathbf{H}_{k1} &
		\mathbf{H}_{k2} &
		\hdots&
		\mathbf{H}_{k\frac{M}{2}} 
	\end{bmatrix} \in \mathbb{C}^{2 \times M} \nonumber \\
	 &\!=\! \begin{bmatrix}
		z_{kV,1V} & \!\!\!z_{kV,1H} &\!\!\! z_{kV,2V} &\!\!\! z_{kV,2H} & \!\!\!\!\hdots\!\!\! &z_{kV,\frac{M}{2}V} & \!\!\!z_{kV,\frac{M}{2}H}\\
		z_{kH,1V} &\!\!\! z_{kH,1H} &\!\!\! z_{kH,2V} &\!\!\! z_{kH,2H} &\!\!\!\! \hdots\!\!\!& z_{kH,\frac{M}{2}V} & \!\!\!z_{kH,\frac{M}{2}H} 
	\end{bmatrix},
\end{align}
where $z_{kX,mY}$ is the channel coefficient between the $X$ polarized component of the $k$th UE's dual-polarized antenna   and $Y$ polarized component of the $m$th  dual-polarized BS antenna  with $m \in \{ 1, \dots, \frac{M}{2} \}$, $k \in \{ 1, \dots, K \}$ and $X, Y \in \left\lbrace V, H \right\rbrace $. Therefore, each block $\mathbf{H}_{km}\in \mathbb{C}^{2 \times 2}$ describes the relation from $V$ to $V$, $V$ to $H$, $H$ to $H$ and $H$ to $V$ polarized waves.

In free-space, the cross-polar transmissions (e.g., from a V polarized BS antenna to a H polarized UE antenna) is zero under ideal conditions. In a practical scenario, the propagation environment causes cross-polarization scattering that changes the initial polarization state of the electromagnetic waves on the way from the transmitter to the receiver. The channel cross-polarization discrimination (XPD) is the channel's ability to maintain radiated or received polarization purity between H and V polarized signals. 
We assume that it is independent of the BS antenna number $m$ and define it for UE $k$ as
\begin{equation}\label{eq3}
	\mathrm{XPD}_k = \frac{ \mathbb{E}\left\lbrace \left| {z}_{kV,mV}\right|^2\right\rbrace }{\mathbb{E}\left\lbrace \left| {z}_{kH,mV}\right|^2\right\rbrace } = \frac{ \mathbb{E}\left\lbrace \left| {z}_{kH,mH}\right|^2\right\rbrace }{\mathbb{E}\left\lbrace \left| {z}_{kV,mH}\right|^2\right\rbrace }=  \frac{1 - q_k}{q_k}
\end{equation}
for a coefficient $ 0\leq q_k \leq 1$. By introducing this coefficient, we obtain
\begin{align}
	&\mathbb{E}\left\lbrace \left| {z}_{kV,mV}\right|^2\right\rbrace  =  \mathbb{E}\left\lbrace \left| {z}_{kH,mH}\right|^2\right\rbrace = \beta_k \left( 1 - q_k\right), \label{eq1}\\
	&\mathbb{E}\left\lbrace \left| {z}_{kH,mV}\right|^2\right\rbrace  =  \mathbb{E}\left\lbrace \left| {z}_{kV,mH}\right|^2\right\rbrace = \beta_k  q_k, \label{eq2}
\end{align}
where $\beta_k$ is the pathloss parameter of UE $k$. Small values of $q_k$ (i.e., high channel XPD) are typically encountered in line-of-sight-dominated outdoor scenarios whereas low channel XPDs are observed in dense scattering environments \cite{degli2011analysis}. Note that \eqref{eq3} only considers the average power values, while the instantaneous ratio between $\left| {z}_{kV,mV}\right|^2$ and $\left| {z}_{kH,mH}\right|^2$ can be as high as $10$ dB due to the polarization selectivity feature of  scattering environments \cite{Asplund2007, oestges2008dual}.

We are considering NLOS communication scenarios that can be modeled by correlated Rayleigh fading. The correlation structure needs to be modeled properly to capture the key properties of both the propagation channel and polarization.
The polarization correlation matrices that define the correlation between the channel coefficients ${z}_{kH,mH}, {z}_{kV,mV}, {z}_{kV,mH}, {z}_{kH,mV}$ can  generally be represented by \cite{coldrey2008modeling}
\begin{equation}
	\mathbf{C}_{\mathrm{BS},k} =\begin{bmatrix}
		1 & t_{k}\\
		t^*_{k} & 1
	\end{bmatrix} \quad \text{and} \quad \mathbf{C}_{\mathrm{UE},k} =\begin{bmatrix}
		1 & r_{k}\\
		r^*_{k} & 1
	\end{bmatrix},
\end{equation}
where the cross-polar correlation (XPC) terms $t_{k}$ and $r_{k}$ at the transmitter and receiver side are defined and computed as
\begin{align}
	&t_{k} = \frac{ \mathbb{E}\left\lbrace {z}_{kV,mV}  {z}^*_{kV,mH}\right\rbrace }{\beta_k\sqrt{q_k\left( 1 - q_k\right)} } = \frac{ \mathbb{E}\left\lbrace {z}_{kH,mV}  {z}^*_{kH,mH}\right\rbrace }{\beta_k\sqrt{q_k\left( 1 - q_k\right)} } ,\\
	&r_{k} = \frac{ \mathbb{E}\left\lbrace {z}_{kV,mV}  {z}^*_{kH,mV}\right\rbrace }{\beta_k\sqrt{q_k\left( 1 - q_k\right)} } = \frac{ \mathbb{E}\left\lbrace {z}_{kH,mH}  {z}^*_{kV,mH}\right\rbrace }{\beta_k\sqrt{q_k\left( 1 - q_k\right)} }. 
\end{align}
Hence, each block in \eqref{eq4} can generally be written as
\begin{equation} \label{eq22:general}
	\mathbf{H}_{km}= \mathbf{\Sigma}_k \odot \left( \mathbf{C}^\frac{1}{2}_{\mathrm{UE},k}\mathbf{G}_{km}  \mathbf{C}^\frac{1}{2}_{\mathrm{BS},k}\right),
\end{equation}
where $\odot$ is the Hadamard (element-wise) product, 
\begin{equation}
	\mathbf{\Sigma}_k = \begin{bmatrix}
		\sqrt{1 - q_k} & \sqrt{ q_k}  \\
		\sqrt{ q_k} & \sqrt{1 - q_k} 
	\end{bmatrix},
\end{equation}\begin{equation}
	  \mathbf{G}_{km} = \begin{bmatrix}
		g_{kV,mV} & g_{kV,mH} \\
		g_{kH,mV} & g_{kH,mH} 
	\end{bmatrix},
\end{equation}
and $\mathbf{G}_{km} $ has i.i.d. circularly symmetric Gaussian entries with $g_{kX,mY} \sim  \mathcal{N}_\mathbb{C}(0, \beta_k)$ for $X,Y \in \left\lbrace V,H\right\rbrace $. 
Various measurements indicate that the transmit and receiver XPCs are close to zero in NLOS scenarios; see \cite[Table 3.1]{clerckx2013mimo}. 
Therefore, we assume that $t_{k}=r_{k}= 0$ when developing the analytical results of this paper. This implies that  the V and H polarized waves fade independently through the channel \cite{Asplund2007}. 
We will study the case when the XPCs are non-zero in Section~\ref{secNumRes}, in which case we will use the expression in \eqref{eq22:general}. By substituting  $\mathbf{C}_{\mathrm{BS},k}  =	\mathbf{C}_{\mathrm{UE},k}  = \mathbf{I}_2$ into \eqref{eq22:general}, we obtain the simplified expression
\begin{equation}\label{eq22}
	\mathbf{H}_{km} = \mathbf{\Sigma}_k \odot \mathbf{G}_{km},
\end{equation}
Eq.~\eqref{eq22} expresses a $2\times2$ Rayleigh fading dual-polarized MIMO channel. Each channel coefficient is scaled by the corresponding XPD coefficient and pathloss  as the propagation environment dictates.

There are multiple dual-polarized antennas at the BS side, thus, the spatial correlation of their fading should also be incorporated into the channel model. If we stack the elements related to UE $k$ for different polarization combinations   as $\mathbf{g}_{k,xy}=[g_{kx,1y}, \dots,g_{kx,\frac{M}{2}y} ] \in \mathbb{C}^{\frac{M}{2} \times 1}$ with $x,y \in \left\lbrace V,H\right\rbrace $, then $\mathbf{g}_{k,xy} \sim \mathcal{N}_\mathbb{C}\left( \mathbf{0}, \mathbf{R}_{\mathrm{BS},k} \right) $ with the spatial correlation matrix $\mathbf{R}_{\mathrm{BS},k} \in \mathbb{C}^{\frac{M}{2} \times \frac{M}{2}}$. 
For example, the vector $\mathbf{g}_{k,VH}$ denotes the relation  (without the XPD coefficients) between the V polarized component of the UE $k$'s antenna and the H polarized component of the BS antennas. Since the V and H polarized antennas are co-located, they see the same scattering environment (i.e., the same scattering objects) and it is, therefore, common to assume equal statistical properties \cite{oestges2008dual,coldrey2008modeling,park2015multi,massivemimobook}.
By following this convention, all the channel vectors have the same spatial correlation matrix $\mathbf{R}_{\mathrm{BS},k}$, i.e.,  $\mathbf{g}_{k,VV} \sim \mathcal{N}_\mathbb{C}\left( \mathbf{0}, \mathbf{R}_{\mathrm{BS},k} \right) $,  $\mathbf{g}_{k,VH} \sim \mathcal{N}_\mathbb{C}\left( \mathbf{0}, \mathbf{R}_{\mathrm{BS},k} \right) $,  $\mathbf{g}_{k,HV} \sim \mathcal{N}_\mathbb{C}\left( \mathbf{0}, \mathbf{R}_{\mathrm{BS},k} \right) $, and  $\mathbf{g}_{k,HH} \sim \mathcal{N}_\mathbb{C}\left( \mathbf{0}, \mathbf{R}_{\mathrm{BS},k} \right) $. In summary, the propagation channel  of UE $k$ becomes
\begin{align}\label{eq6}
	\mathbf{H}_k &= \begin{bmatrix}
		\mathbf{H}_{k1} &
		\mathbf{H}_{k2} &
		\hdots&
		\mathbf{H}_{k\frac{M}{2}} 
	\end{bmatrix} \nonumber \\
	&= \left( \mathbf{1}_{1 \times \frac{M}{2}} \otimes \mathbf{\Sigma}_k  \right) \odot \begin{bmatrix}
		\mathbf{G}_{k1} &
		\mathbf{G}_{k2} &
		\hdots&
		\mathbf{G}_{k\frac{M}{2}} 
	\end{bmatrix} 
	\nonumber \\
	&=\left( \mathbf{1}_{1 \times \frac{M}{2}} \otimes \mathbf{\Sigma}_k  \right) \odot \left(   \mathbf{S}_k \left( \mathbf{R}_{\mathrm{BS},k} \otimes \mathbf{I}_{2}\right)^{1/2}  \right), 
\end{align}
where  
\begin{align}
\mathbf{S}_k &=\begin{bmatrix}
	s_{kV,1V} &\!\!\!\! s_{kV,1H} &\!\! \!\!\!\hdots &\!\!\!\! s_{kV,\frac{M}{2}V} &\!\!\!\! s_{kV,\frac{M}{2}H}\\
	s_{kH,1V} & \!\!\! s_{kH,1H} &\!\! \!\!\! \hdots&\!\!\!\! s_{kH,\frac{M}{2}V} &\!\!\!\! s_{kH,\frac{M}{2}H} 
\end{bmatrix} \nonumber \\
&= \begin{bmatrix} \mathbf{s}_{kV}^H \\ \mathbf{s}_{kH}^H\end{bmatrix}  
\in \mathbb{C}^{2 \times M}
\end{align}
has i.i.d. entries with $\mathcal{N}_\mathbb{C}(0, 1 )$-distribution and the operator $\otimes$ denotes the Kronecker product.
The vectors $\mathbf{s}_{kV},\mathbf{s}_{kH} \in \mathbb{C}^{M\times 1}$ gathers the elements on the first and second line of $\mathbf{S}_k$.

The channel matrix $\mathbf{H}_k = [
	\mathbf{h}_{kV} \quad  \mathbf{h}_{kH}
]^H \in \mathbb{C}^{2 \times M}$ contains the rows $\mathbf{h}_{kV},\mathbf{h}_{kH} \in \mathbb{C}^{M\times 1}$. 
By introducing the notation $\mathbf{R}_k = \mathbf{R}_{\mathrm{BS},k}  \otimes \mathbf{I}_2$, we can express these rows as
\begin{align}
	\mathbf{h}_{kV} &=  \left( \mathbf{1}_{1 \times \frac{M}{2}} \otimes [  \sqrt{1-q_k} \quad \sqrt{q_k}]   \right) \, \odot \, \mathbf{R}^{1/2}_{k} \mathbf{s}_{kV} \nonumber \\ & = \left( \mathbf{I}_{\frac{M}{2}} \otimes \begin{bmatrix} \sqrt{1-q_k} & 0 \\ 0 & \sqrt{q_k} \end{bmatrix}   \right) \mathbf{R}^{1/2}_{k} \mathbf{s}_{kV} 
	\end{align}
	and
\begin{align}
	\mathbf{h}_{kH} &= \left( \mathbf{1}_{1 \times \frac{M}{2}} \otimes [  \sqrt{q_k} \quad \sqrt{1-q_k}]   \right) \, \odot \, \mathbf{R}^{1/2}_{k} \mathbf{s}_{kV} \nonumber \\ &
	= \left( \mathbf{I}_{\frac{M}{2}} \otimes \begin{bmatrix} \sqrt{q_k} & 0 \\ 0 & \sqrt{1-q_k} \end{bmatrix}   \right) \mathbf{R}^{1/2}_{k} \mathbf{s}_{kH} .
\end{align}

The covariance matrices of these vectors will be utilized during the channel estimation and are computed as
\begin{align}
	\mathbb{E}\left\lbrace \mathbf{h}_{kV} \mathbf{h}^H_{kV}\right\rbrace 
	&= \mathbf{R}^{1/2}_{k} 
	 \left( \mathbf{I}_{\frac{M}{2}} \otimes \begin{bmatrix} 1-q_k & 0 \\ 0 & q_k \end{bmatrix}   \right)
	 	 \left( \mathbf{R}^{1/2}_{k}\right)^H \nonumber \\
 &\triangleq  \mathbf{R}_{kV}, \label{eq:Rvk}
\end{align}
and 
\begin{align}
	\mathbb{E}\left\lbrace \mathbf{h}_{kH} \mathbf{h}^H_{kH}\right\rbrace  
	&= \mathbf{R}^{1/2}_{k}
	\left( \mathbf{I}_{\frac{M}{2}} \otimes \begin{bmatrix} q_k & 0 \\ 0 & 1-q_k \end{bmatrix}   \right) 
	 \left( \mathbf{R}^{1/2}_{k}\right)^H \nonumber \\
	 & \triangleq \mathbf{R}_{kH}, \label{eq:Rhk}
\end{align}
where $\mathbf{R}_{kV} + \mathbf{R}_{kH} = \mathbf{R}_{k}$. Using this new notation, we can also express the channel matrix as $	\mathbf{H}_k  = \begin{bmatrix}
	\mathbf{R}_{kV}^{1/2} \mathbf{s}_{kV} \quad
		\mathbf{R}_{kH}^{1/2} \mathbf{s}_{kH}
	\end{bmatrix}^H$.
Moreover, we can calculate the covariance matrix of the entire channel matrix as
\begin{align}\label{eq5}
	\mathbb{E}\left\lbrace \mathrm{vec}\left(\mathbf{H}^H_k \right)\mathrm{vec}\left(\mathbf{H}^H_k \right)^H\right\rbrace & = \begin{bmatrix}
		\mathbf{R}_{kV} & \mathbf{0}\\
		\mathbf{0} &  \mathbf{R}_{kH}
	\end{bmatrix} \nonumber \\
&\triangleq\mathbf{\Delta}_k \! \in \mathbb{C}^{2M \times 2M},
\end{align} 
where $\mathrm{vec}(\cdot)$ denotes vectorization. Notice that \eqref{eq5} implies $	\mathbb{E}\left\lbrace \mathbf{h}_{kH} \mathbf{h}^H_{kV}\right\rbrace = 	\mathbb{E}\left\lbrace \mathbf{h}_{kV} \mathbf{h}^H_{kH}\right\rbrace = \mathbf{0}$
 since the  V and H polarized waves fade independently through the channel.
 
 \begin{remark} 
 The channel model can be enriched to also capture hardware polarization effects; in particular, a limited cross-polar isolation (XPI) so that the signal meant for one polarization leaks into the opposite polarization.
 For well-designed antennas, the antenna depolarization effects can be made negligible (e.g., XPIs on the order of 30 dBs or more) in contrast to the propagation channel depolarization that is given by the environment \cite[Ch. 8]{stutzman2018polarization}. If the XPI is estimated, it can also be inverted in the digital baseband. Hence, we are not considering XPI in this paper to keep the notation relatively simple.
 \end{remark}

\section{Channel Estimation}
\label{secCE}
Each BS requires CSI for uplink receive processing and downlink transmit precoding. Therefore, $\tau_p$ samples are reserved for performing uplink pilot-based channel estimation in each coherence block, giving room for $\tau_p$ mutually orthogonal pilot sequences. Following \cite{Li2016b,Bjornson2010a}, each UE sends its pilot signal $\boldsymbol{\Phi}_k \in \mathbb{C}^{2\times \tau_p}$ to the BS with $\tau_p =2K$ (and $2K \leq \tau_c$) to estimate all channel dimensions at the BS. 
 The pilot signal is designed as $\boldsymbol{\Phi}_k = \mathbf{L}^{1/2}_k \mathbf{V}^T_k$ where $\mathbf{L}_k = \mathrm{diag}\left( p_{kV}, p_{kH}\right) $ is a pilot allocation matrix with $p_{kV}, p_{kH}$ being the pilot powers  allocated to the V and H polarized antennas, respectively.
 The orthogonal pilot matrix $\mathbf{V}_k \in \mathbb{C}^{ \tau_p \times 2}$ is designed so that $\mathbf{V}^H_k \mathbf{V}_k = \tau_p\mathbf{I}_{2} $ and $\mathbf{V}^H_k \mathbf{V}_l = \mathbf{0}_2 $ if $l \neq k$.  Also, $\mathrm{tr}\left(\boldsymbol{\Phi}_k \boldsymbol{\Phi}^H_k  \right)/ \tau_p \leq P_k $ where $P_k$ is the total uplink pilot power of UE $k$. Thus, 
\begin{equation}
	\boldsymbol{\Phi}_k\mathbf{V}^*_k =  \begin{bmatrix}
		\sqrt{p_{kV}}\tau_p,  & 0\\
		0 & \sqrt{p_{kH}}\tau_p
	\end{bmatrix} =  \tau_p\mathbf{L}^{1/2}_k,
\end{equation}
\begin{equation}
	\boldsymbol{\Phi}_k\mathbf{V}^*_l  = \mathbf{0}_{2}, \quad l \neq k.
\end{equation}

All UEs transmit their pilot signals simultaneously. 
The received pilot signal  $\mathbf{Y}   \in  \mathbb{C}^{M \times \tau_p}$ at the BS is then given by 
\begin{equation}
	\mathbf{Y} = \sum_{l=1}^K \mathbf{H}^H_l \boldsymbol{\Phi}_l +\mathbf{N},
\end{equation}
where $\mathrm{vec}\left(\mathbf{N} \right) \sim \mathcal{N}_\mathbb{C}( \mathbf{0}, \sigma^2_\mathrm{ul} \mathbf{I}_{M\tau_p} ) $ is the receiver noise with variance $\sigma^2_\mathrm{ul}$. To estimate the channel of UE $k$, the BS can first process the receive signal by correlating it with the UE's pilot signal.
The processed pilot signal $ \mathbf{Y}^p_k  \in \mathbb{C}^{M \times 2}$ is
\begin{equation}\label{eq_processed}
	\mathbf{Y}^p_k =	\mathbf{Y}\mathbf{V}^*_k = \tau_p\mathbf{H}^H_k  \mathbf{L}^{1/2}_k+ \mathbf{N}\mathbf{V}^*_k .
\end{equation}
Vectorizing  \eqref{eq_processed} gives
\begin{equation}\label{eqrec}
	\mathrm{vec}\left(\mathbf{Y}^p_k \right) =\mathbf{A}  \mathrm{vec}\left( \mathbf{H}^H_k \right) + \mathrm{vec}\left(\mathbf{N}\mathbf{V}^*_k \right),
\end{equation}
where  $\mathbf{A} = \left( \tau_p\mathbf{L}^{1/2}_k \otimes \mathbf{I}_{M}\right) $ and $\mathrm{vec}\left(\mathbf{N}\mathbf{V}^*_k \right)  \sim \mathcal{N}_\mathbb{C}\left( \mathbf{0}, \sigma^2_\mathrm{ul} \tau_p\mathbf{I}_{2M} \right)$. Besides, the received processed pilot signal can be written as $\mathrm{vec}\left(\mathbf{Y}^p_k \right) \triangleq \begin{bmatrix}
	\mathbf{y}^p_{kV}\\
	\mathbf{y}^p_{kH}
\end{bmatrix} $ where
\begin{align}
\mathbb{E}\left\lbrace\mathbf{y}^p_{kV} (\mathbf{y}^p_{kV})^H\right\rbrace & = 
	\tau_p\left(p_{kV}\tau_p\mathbf{R}_{kV} + \sigma^2_\mathrm{ul}\mathbf{I}_M \right) \nonumber \\
	&\triangleq\tau_p \left( \mathbf{\Psi}^v_k\right)^{-1},
\end{align}
\begin{align}
	\mathbb{E}\left\lbrace\mathbf{y}^p_{kH} (\mathbf{y}^p_{kH})^H\right\rbrace & = 
	\tau_p\left(p_{kH}\tau_p\mathbf{R}_{kH} + \sigma^2_\mathrm{ul}\mathbf{I}_M \right) \nonumber \\
	&\triangleq \tau_p \left( \mathbf{\Psi}^h_k\right)^{-1} .
\end{align}

Then, based on \eqref{eqrec}, the MMSE estimate of $\mathbf{H}_k $ is \cite{Bjornson2010a}
\begin{align}
	\mathrm{vec}\left( \hat{\mathbf{H}}^H_k \right) &= \mathbf{\Delta}_k\mathbf{A}^H \left(\mathbf{A}\mathbf{\Delta}_k \mathbf{A}^H + \sigma^2_\mathrm{ul} \tau_p\mathbf{I}_{2M} \right)^{-1} \mathrm{vec}\left(\mathbf{Y}^p_k \right)\nonumber \\
	&=\begin{bmatrix}
		\sqrt{p_{kV}}\mathbf{R}_{kV} \mathbf{\Psi}^v_k 	\mathbf{y}^p_{kV}\\
		\sqrt{p_{kH}}\mathbf{R}_{kH}\mathbf{\Psi}^h_k 	\mathbf{y}^p_{kH}
	\end{bmatrix} \triangleq \begin{bmatrix}
		\hat{\mathbf{h}}_{kV} \\\hat{\mathbf{h}}_{kH}
	\end{bmatrix},
\end{align}
where the MMSE estimates associated with V/H antennas are uncorrelated random variables:
\begin{align}
&\hat{\mathbf{h}}_{kV} \sim  \mathcal{N}_\mathbb{C}\left( \mathbf{0}, \mathbf{\Gamma}_k^v\right), \\
&\hat{\mathbf{h}}_{kH} \sim  \mathcal{N}_\mathbb{C}\left( \mathbf{0}, \mathbf{\Gamma}_k^h\right),
\end{align}
with $\mathbf{\Gamma}_k^v =	{p_{kV}}\tau_p \mathbf{R}_{kV} \mathbf{\Psi}^v_k 	\mathbf{R}_{kV}$ and $\mathbf{\Gamma}_k^h =	{p_{kH}}\tau_p \mathbf{R}_{kH} \mathbf{\Psi}^h_k 	\mathbf{R}_{kH}$ where $\mathrm{tr}\left( \mathbf{\Gamma}_k^v\right) = \mathrm{tr}\left( \mathbf{\Gamma}_k^h\right)$ for equal pilot powers ${p_{kV}} = {p_{kH}}$. Note that the estimates $\hat{\mathbf{h}}_{kV} $ and $\hat{\mathbf{h}}_{kH}$ are uncorrelated since the channels $\mathbf{h}_{kV}$ and $\mathbf{h}_{kH}$ are uncorrelated.
The error covariance matrix is
\begin{align}
	\mathbf{C}_{\mathrm{MMSE},k} &= \mathbf{\Delta}_k -\mathbf{\Delta}_k\mathbf{A}^H \left(\mathbf{A}\mathbf{\Delta}_k \mathbf{A}^H + \sigma^2_\mathrm{ul} \tau_p\mathbf{I}_{2M} \right)^{-1} \mathbf{A} \mathbf{\Delta}_k \nonumber\\
	&=\begin{bmatrix}
		\mathbf{R}_{kV} -\mathbf{\Gamma}_k^v& \mathbf{0}\\
		\mathbf{0}&	\mathbf{R}_{kH}- \mathbf{\Gamma}_k^h
	\end{bmatrix} \! \triangleq \! \begin{bmatrix}
		\mathbf{C}^{v}_{k} & \mathbf{0}\\
		\mathbf{0} & \mathbf{C}^{h}_{k}
	\end{bmatrix}\!.
\end{align}
These results will be utilized in the uplink and downlink transmissions to design combining and precoding schemes.

\section{Uplink Transmission}

 In the uplink data transmission phase, the received signal at the BS is
\begin{align}\label{eqULreceived}
	\mathbf{y} &= \sum_{l=1}^K \mathbf{H}^H_l \mathbf{P}^{1/2}_l  \mathbf{x}_l  + \mathbf{n},  
\end{align}
where $ \mathbf{P}_l =\begin{bmatrix}
	{\rho^\mathrm{ul}_{lV}} & 0 \\ 0 & 	{\rho^\mathrm{ul}_{lH}}
\end{bmatrix}$ is the uplink transmit power allocation matrix, $ \mathbf{x}_l = \begin{bmatrix}
x_{lV}\\ x_{lH}
\end{bmatrix} \sim \mathcal{N}_\mathbb{C}(\mathbf{0}, \mathbf{I}_2)$ is the data signals and $\mathbf{n} \sim  \mathcal{N}_\mathbb{C}(\mathbf{0}, \sigma^2_\mathrm{ul}\mathbf{I}_M) $ is the receiver noise. According to the block-fading assumption, at the beginning of each coherence block, the channels realizations are unknown. Then, in the channel estimation phase of the TDD protocol, the channels are estimated at the BS side and the UEs do not have instantaneous CSIs. Therefore, the precoder matrix of user $k$ is selected as an identity matrix  in \eqref{eqULreceived} to send one signal per polarization. 
Note that the uplink of a frequency-division duplex (FDD) system can be implemented identically to the uplink of a TDD system, thus the expressions derived in this section are also applicable in that case.

\subsection{Uplink Linear Detection}
First, we consider the case in which the data signals from each UE antenna, $x_{lV} $ and $x_{lH}$, are decoded simultaneously while treating the other streams as noise.  A linear detector $\mathbf{v}_{ki} \in \mathbb{C}^{M \times 1} $  for $i \in \left\lbrace V,H\right\rbrace $ based on the channel estimates is applied to the received signal as
\begin{align}\label{eqUL}
	\mathbf{v}^H_{ki} \mathbf{y} &=  \sum_{l=1}^K \sqrt{\rho^\mathrm{ul}_{lV}} \mathbf{v}^H_{ki}\mathbf{h}_{lV} x_{lV} +\sqrt{\rho^\mathrm{ul}_{lH}} \mathbf{v}^H_{ki}\mathbf{h}_{lH} x_{lH} \nonumber \\
	 &+ \mathbf{v}^H_{ki}\mathbf{n}.
\end{align}
Then, we can rewrite \eqref{eqUL} as
\begin{align}
	\mathbf{v}^H_{ki} \mathbf{y}  &=  \sqrt{\rho^\mathrm{ul}_{ki}} \mathbb{E}\left\lbrace  \mathbf{v}^H_{ki}\mathbf{h}_{ki}\right\rbrace  x_{ki} \nonumber \\
	 &+ \sqrt{\rho^\mathrm{ul}_{ki}} \left( \mathbf{v}^H_{ki}\mathbf{h}_{ki} -  \mathbb{E}\left\lbrace  \mathbf{v}^H_{ki}\mathbf{h}_{ki}\right\rbrace \right)  x_{ki} \nonumber \\  
	   &+\sqrt{\rho^\mathrm{ul}_{ki'}}  \mathbf{v}^H_{ki'}\mathbf{h}_{ki'} x_{ki'}  +\sum_{l \neq k} \sqrt{\rho^\mathrm{ul}_{lV}} \mathbf{v}^H_{ki}\mathbf{h}_{lV} x_{lV} \nonumber \\
	   & +\sqrt{\rho^\mathrm{ul}_{lH}} \mathbf{v}^H_{ki}\mathbf{h}_{lH} x_{lH} + \mathbf{v}^H_{ki}\mathbf{n},
\end{align}
by adding and subtracting the averaged precoded channel $\mathbb{E}\left\lbrace  \mathbf{v}^H_{ki}\mathbf{h}_{ki}\right\rbrace$. Let $i'$ denote the opposite polarization so that $i \cup i' = \left\lbrace V,H\right\rbrace $ and $i \neq  i'$. The desired signal received over the averaged precoded channel $\mathbb{E}\left\lbrace  \mathbf{v}^H_{ki}\mathbf{h}_{ki}\right\rbrace$ is treated as the true desired signal. The  part received  over $\left(  \mathbf{v}^H_{ki}\mathbf{h}_{ki} -  \mathbb{E}\left\lbrace  \mathbf{v}^H_{ki}\mathbf{h}_{ki}\right\rbrace \right) $ is treated as uncorrelated noise. The following lemma provides a lower bound on the uplink capacity, which makes it an achievable SE. This bound is referred to as the use-and-then-forget (UaTF) \cite{massivemimobook} since the channel estimates are utilized in the receiver combining and then forgotten before the signal detection.

\begin{lemma}
The uplink achievable SE of UE $k$  using  the UaTF  bound is
\begin{equation}\label{eqUaTF}
	R^\mathrm{ul}_k  = \frac{\tau_c -\tau_p}{\tau_c} \sum_{i \in \left\lbrace V,H\right\rbrace }  \log_2\left( 1 +\gamma_{ki} \right)
\end{equation}
with  $\gamma_{ki}$ is given in \eqref{floatFig1}, at the top of next page.
\begin{figure*}
\begin{align}\label{floatFig1}
	\gamma_{ki} = 
	\frac{\rho^\mathrm{ul}_{ki} \left|\mathbb{E}\left\lbrace  \mathbf{v}^H_{ki}\mathbf{h}_{ki}\right\rbrace\right|^2   }{ \displaystyle\sum_{l=1}^K \rho^\mathrm{ul}_{lV} \mathbb{E}\left\lbrace\left|  \mathbf{v}^H_{ki}\mathbf{h}_{lV}\right|^2 \right\rbrace + \rho^\mathrm{ul}_{lH} \mathbb{E}\left\lbrace \left| \mathbf{v}^H_{ki}\mathbf{h}_{lH}\right|^2 \right\rbrace -\rho^\mathrm{ul}_{ki} \left|\mathbb{E}\left\lbrace  \mathbf{v}^H_{ki}\mathbf{h}_{ki}\right\rbrace\right|^2   + \sigma^2_\mathrm{ul} \mathbb{E}\left\lbrace \left\|  \mathbf{v}_{ki}\right\| ^2 \right\rbrace}.
\end{align}
\end{figure*}

\end{lemma}
 \begin{IEEEproof}
 	The derivation is similar to that in \cite[Theorem 4.4]{massivemimobook} and is therefore  omitted.
 \end{IEEEproof}

The uplink achievable SE in \eqref{eqUaTF} can be computed for any choice of combining vector. The MMSE detector is
\begin{align}\label{vMMMSE}
	\mathbf{v}^\mathrm{MMSE}_{ki} =\sqrt{\rho^\mathrm{ul}_{ki}} \mathbf{\Upsilon}^{-1} \hat{\mathbf{h}}_{ki},
\end{align}
where $\mathbf{\Upsilon} = \sum_{l=1}^K \hat{\mathbf{H}}^H_l \mathbf{P}_l \hat{\mathbf{H}}_l + \rho^\mathrm{ul}_{lV} \mathbf{C}^v_l +\rho^\mathrm{ul}_{lH} \mathbf{C}^h_l + \sigma^2_\mathrm{ul} \mathbf{I}_{M} $.
Other potential selections of $\mathbf{v}_{ki}$ for low complexity can be  ZF  and MR combining vectors as
\begin{align}
	&\mathbf{v}^\mathrm{ZF}_{ki} = \left[\hat{\mathbf{H}}_\mathrm{all} \left( \hat{\mathbf{H}}^H_\mathrm{all} \hat{\mathbf{H}}_\mathrm{all} \right)^{-1} \right]_{ki}, \label{vZF} \\
	&\mathbf{v}^\mathrm{MR}_{ki} = \hat{\mathbf{h}}_{ki},\label{vMR}
\end{align}
where $\hat{\mathbf{H}}_\mathrm{all}= \left[ \hat{\mathbf{H}}^H_1, \dots, \hat{\mathbf{H}}^H_K\right] =\left[\hat{\mathbf{h}}_{1V} \ \hat{\mathbf{h}}_{1H}, \dots,\hat{\mathbf{h}}_{KV}  \ \hat{\mathbf{h}}_{KH} \right]  \in \mathbb{C}^{M \times 2K}  $ and $\left[ . \right]_{ki}$ denotes the ${ki}^{th}$ column  corresponds to $k^{th}$ UE and $i \in \left\lbrace V,H\right\rbrace $. It is expected that MR combining will provide lower SEs than the other combining vectors but it does not require any matrix inversion. In the following lemma,  \eqref{eqUaTF}  is computed in closed form for MR combining.
\begin{lemma}
	If MR combining $\mathbf{v}^\mathrm{MR}_{ki} = \hat{\mathbf{h}}_{ki}$ is used based on the MMSE estimator, then the achievable SE in \eqref{eqUaTF} can be computed in closed form  as given in \eqref{eqRateUL}, at the top of next page.
	
	\begin{figure*}
\begin{align}\label{eqRateUL}
	\!\!\!\!\!\!\! R^\mathrm{ul}_k =  \frac{\tau_c -\tau_p}{\tau_c} \left[ \log_2\left(1 +\frac{	\rho^\mathrm{ul}_{kV}  \mathrm{tr}\left(\mathbf{\Gamma}_k^v\right) }{\sum_{l=1}^K \left(  \rho^\mathrm{ul}_{lV}  \frac{\mathrm{tr}\left( \mathbf{\Gamma}_k^v  \mathbf{R}^{v}_{l}\right)}{\mathrm{tr}\left( \mathbf{\Gamma}_k^v \right)} +\rho^\mathrm{ul}_{lH} \frac{\mathrm{tr}\left( \mathbf{\Gamma}_k^v  \mathbf{R}^{h}_{l}\right)}{\mathrm{tr}\left( \mathbf{\Gamma}_k^v \right)}\right)  + \sigma^2_\mathrm{ul}}\right) 
	+ \log_2\left(1 + \frac{	\rho^\mathrm{ul}_{kH} \mathrm{tr}\left(\mathbf{\Gamma}_k^h\right) }{\sum_{l=1}^K \left( \rho^\mathrm{ul}_{lH} \frac{\mathrm{tr}\left( \mathbf{\Gamma}_k^h  \mathbf{R}^{h}_{l}\right)}{\mathrm{tr}\left( \mathbf{\Gamma}_k^h \right)} +\rho^\mathrm{ul}_{lV} \frac{\mathrm{tr}\left( \mathbf{\Gamma}_k^h  \mathbf{R}^{v}_{l}\right)}{\mathrm{tr}\left( \mathbf{\Gamma}_k^h \right)} \right) + \sigma^2_\mathrm{ul}} \right)\right]  .
\end{align}
\vspace{-0.8cm}
	\end{figure*}
	
\end{lemma}
\begin{IEEEproof}
	Similar to Appendix B.
\end{IEEEproof}

The closed-form expression in \eqref{eqRateUL} provides insights into the basic behaviors of dual-polarized massive MIMO. The first and second logarithms represent the SE of the data streams associated with V/H polarizations, respectively. The signal terms in the numerators are proportional to the total variances 
$\mathrm{tr}\left(\mathbf{\Gamma}_k^v\right)$ and $\mathrm{tr}\left(\mathbf{\Gamma}_k^h\right)$ of the channel estimates, thus, a beamforming gain proportional to $M$ is obtained (similar to uni-polarized massive MIMO). The denominators contain interference terms from both polarizations of all UEs. The two terms have a similar form but different sizes depending on how $q_k$ enters into the expressions. Since the interference terms are ratios of traces, they will not grow with the number of antennas. 

The simplified version of \eqref{eqRateUL} for spatially uncorrelated channels, $\mathbf{R}_{\mathrm{BS},k} = \beta_k \mathbf{I}_{\frac{M}{2}}$ for $k=1,\dots,K$, is given in \eqref{eqrateULSimplified}, at the top of next page, where $
\gamma_{kV,1} = \frac{ p_{kV} \tau_p \beta^2_k (1-q_k)^2}{ p_{kV} \tau_p \beta_k (1 - q_k ) + \sigma_\mathrm{ul}^2} $, $\gamma_{kV,2} =  \frac{  p_{kV} \tau_p \beta^2_k q_k^2}{ p_{kV} \tau_p \beta_k  q_k  + \sigma_\mathrm{ul}^2}$, $\gamma_{kH,1}=  \frac{  p_{kH} \tau_p \beta^2_k (1-q_k)^2}{ p_{kH} \tau_p \beta_k (1 - q_k ) + \sigma_\mathrm{ul}^2}$ and $
\gamma_{kH,2}= \frac{ p_{kH} \tau_p \beta^2_k q_k^2}{ p_{kH} \tau_p \beta_k  q_k  + \sigma_\mathrm{ul}^2}$. Notice that there is a beamforming gain of $\frac{M}{2}$ for each data stream, which is proportional to the total number of antennas, but equal to the number of antennas per polarization. Besides, the interference terms are subject to non-coherent combining since they do not scale with  the number of antennas and are products of the data transmission powers and channel gains. This is aligned with the previous discussion regarding \eqref{eqRateUL} but seen more clearly in \eqref{eqrateULSimplified}.
\begin{figure*}
\begin{align}\label{eqrateULSimplified}
	\!\!\!\!\!\!	R^\mathrm{ul}_k&=\frac{\tau_c -\tau_p}{\tau_c}\log_2\left(1 +\frac{\frac{M}{2}\rho_{kV}^\mathrm{ul}(\gamma_{kV,1}   + \gamma_{kV,2} ) }{\displaystyle \sum_{l=1}^K \left(  \rho^\mathrm{ul}_{lV} \frac{\gamma_{kV,1} \beta_l (1 -q_l)   + \gamma_{kV,2} \beta_l q_l}{\gamma_{kV,1}   + \gamma_{kV,2} } + \rho^\mathrm{ul}_{lH} \frac{\gamma_{kV,1} \beta_l q_l  + \gamma_{kV,2} \beta_l (1 -q_l) }{\gamma_{kV,1}  + \gamma_{kV,2}  }\right)   + \sigma_\mathrm{ul}^2  }\right) \nonumber \\
	&+ \frac{\tau_c -\tau_p}{\tau_c} \log_2\left(1 + \frac{\frac{M}{2}\rho_{kH}^\mathrm{ul}(\gamma_{kH,1}   + \gamma_{kH,2} ) }{\displaystyle \sum_{l=1}^K \left(  \rho^\mathrm{ul}_{lH} \frac{\gamma_{kH,1} \beta_l (1 -q_l)   + \gamma_{kH,2} \beta_l q_l}{\gamma_{kH,1}   + \gamma_{kH,2} } + \rho^\mathrm{ul}_{lV} \frac{\gamma_{kH,1} \beta_l q_l  + \gamma_{kH,2} \beta_l (1 -q_l) }{\gamma_{kH,1}  + \gamma_{kH,2}  }\right)   + \sigma_\mathrm{ul}^2  } \right), 
\end{align}
\hrule
\end{figure*}

\subsection{Uplink MMSE-SIC detection}

Using the linear detection method above, we observe that the self-interference that is caused by the data stream corresponding to the opposite polarization at the same UE is not suppressed. Alternatively, the UE may apply an MMSE successive interference cancellation (SIC) detector to detect the streams since the BS has the estimated channels $\hat{\mathbf{H}}_1,  \dots, \hat{\mathbf{H}}_K$.  The lower bound on the uplink SE when the MMSE-SIC scheme is used is given in the following lemma.  

\begin{lemma}
	\label{lemmaULSIC}
	Using per-stream MMSE-SIC decoding at the BS, the achievable SE of UE $k$ using MMSE-SIC is 
	\begin{align} 
		R^\mathrm{ul,SIC}_k 
		&=\frac{\tau_c -\tau_p}{\tau_c}\mathbb{E}\left\lbrace \! \log_2 \mathrm{det} \!\! \left(  \!\mathbf{I}_2 + \mathbf{P}_{k}\hat{\mathbf{H}}_{k}\! \left( \sum_{l = k+1}^K   {\rho^\mathrm{ul}_{lV}} \hat{\mathbf{h}}_{lV} \hat{\mathbf{h}}^H_{lV} \right. \right.\right. \nonumber \\
		& \left. \left.\left.+ {\rho^\mathrm{ul}_{lH}} \hat{\mathbf{h}}_{lH} \hat{\mathbf{h}}^H_{lH}+ \sum_{l = 1}^K     \rho^\mathrm{ul}_{lV} \mathbf{C}^v_l +\rho^\mathrm{ul}_{lH} \mathbf{C}^h_l + \sigma^2_\mathrm{ul} \mathbf{I}_{M}\right)^{-1}\!\!\!\!\!\! \hat{\mathbf{H}}^H_{k}  \right) \!\!	\right\rbrace,
	\end{align}
	and the achievable uplink sum SE is
	\begin{align}
		R^\mathrm{ul,SIC} &= \sum_{l = 1}^K R^\mathrm{ul,SIC}_l \nonumber \\
		 &= \frac{\tau_c -\tau_p}{\tau_c} \mathbb{E}\left\lbrace \! \log_2 \mathrm{det} \!\! \left( \! \mathbf{I}_M + \sum_{l = 1}^K\hat{\mathbf{H}}^H_{l} \mathbf{P}_{l} \hat{\mathbf{H}}_{l} \right.\right.   \nonumber\\
		 &\left.\left. \times \left(  \sum_{j = 1}^K     \rho^\mathrm{ul}_{jV} \mathbf{C}^v_j +\rho^\mathrm{ul}_{jH} \mathbf{C}^h_j + \sigma^2_\mathrm{ul} \mathbf{I}_{M}\!\right)^{-1}	\right) \!\! \right\rbrace.
	\end{align}
\end{lemma}
\begin{IEEEproof}
	The proof is given in Appendix A.
\end{IEEEproof}
The MMSE-SIC procedure may be a computationally heavy process depending on the number of streams since the signals  need to be buffered.

\section{Downlink Transmission}

In the downlink data transmission, the BS transmits simultaneously to all UEs using precoding computed based on the channel estimates derived in Section  \ref{secCE}. The transmitted downlink signal is 
\begin{equation}
	\mathbf{x} = \sum_{l=1}^K \mathbf{W}_l \mathbf{d}_l,
\end{equation}
where   $\mathbf{d}_l = [d_{lV} \,\, d_{lH}]^T \in \mathbb{C}^{2 \times 1}$ is the transmit signal satisfying $\mathbb{E}\left\lbrace\mathbf{d}_l \mathbf{d}^H_l \right\rbrace = \mathbf{I}_2 $ and  $\mathbf{W}_l = [\mathbf{w}_{lV} \,\, \mathbf{w}_{lH} ] \in \mathbb{C}^{M \times 2}$ is the downlink precoding matrix such that $\mathrm{tr}\left( \mathbb{E}\left\lbrace \mathbf{W}^H_l\mathbf{W}_l \right\rbrace\right) \leq \rho^\mathrm{dl}_{lV} + \rho^\mathrm{dl}_{lH} $ where  the transmit powers of V/H the antennas are denoted $\rho^\mathrm{dl}_{lV}$ and $\rho^\mathrm{dl}_{lH}$, respectively.

The received signal at UE $k$ is denoted by $\mathbf{y}_k \in \mathbb{C}^{2 \times 1}$ and computed as 
\begin{align}\label{eqRecDL}
	\mathbf{y}_k &= \mathbf{H}_k \mathbf{x} + \mathbf{n}_k = \mathbf{H}_k \mathbf{W}_k \mathbf{d}_k +\mathbf{H}_k \sum_{\substack{l=1 \\ l\neq k}}^K \mathbf{W}_l \mathbf{d}_l + \mathbf{n}_k ,
\end{align}
where $\mathbf{n}_k \sim \mathcal{N}_\mathbb{C}\left( \mathbf{0}, \sigma^2_\mathrm{dl}\mathbf{I}_2\right) $ is the receiver noise.  The first term in \eqref{eqRecDL} corresponds to the desired signal whereas the second term is the interference caused by transmissions to other users.

\subsection{Downlink Linear Processing}
 The UEs do not have instantaneous CSIs since no downlink pilots are sent.  However, their average effective channels $\mathbb{E}\left\lbrace  \mathbf{H}_k \mathbf{W}_k  \right\rbrace$ are known. The UEs can detect each data symbol separately using the linear MMSE combining vector $\mathbf{v}_{\mathrm{dl},ki}  \in \mathbb{C}^{2\times 1}$ as \cite{Li2016b} 
\begin{align} \label{vMMSE}
	 \mathbf{v}_{\mathrm{dl},ki} & =  (\mathbb{E}\{ \mathbf{y}_k \mathbf{y}_k^{H} \} )^{-1}  \mathbb{E}\left\lbrace  \mathbf{H}_k \mathbf{w}_{ki}  \right\rbrace  \nonumber \\
	 &=  \left(  \mathbb{E}\left\lbrace  \mathbf{H}_k \sum_{l=1}^K \mathbf{W}_l \mathbf{W}^H_l  \mathbf{H}^H_k \right\rbrace + \sigma^2_\mathrm{dl} \mathbf{I}_2    \right)^{-1}  \mathbb{E}\left\lbrace  \mathbf{H}_k \mathbf{w}_{ki}  \right\rbrace 
\end{align}
for $i \in \left\lbrace V, H\right\rbrace $.
By applying $\mathbf{v}_{\mathrm{dl},ki} $ to the received signal in \eqref{eqRecDL}, we can obtain the following.

\begin{lemma}
An achievable downlink SE is
\begin{equation} \label{eq:Rate-woSIC}
	R^\mathrm{woSIC}_{k} = \frac{\tau_c -\tau_p}{\tau_c} \sum_{i \in \left\lbrace V,H\right\rbrace }\log_2\left( 1 + \eta^\mathrm{dl}_{ki}\right), 
\end{equation}
where the SINR $\eta^\mathrm{dl}_{ki}$ is given in \eqref{floatFig2}, at the top of next page. 
\begin{figure*}
	
\begin{equation} \label{floatFig2}
	\eta^\mathrm{dl}_{ki} = \frac{\left|\mathbf{v}^H_{\mathrm{dl},ki} \mathbb{E}\left\lbrace  \mathbf{H}_k \mathbf{w}_{ki}  \right\rbrace \right|^2 }{\mathbf{v}^H_{\mathrm{dl},ki}\left(   \mathbb{E}\left\lbrace  \mathbf{H}_k \sum_{l=1}^K \mathbf{W}_l \mathbf{W}^H_l  \mathbf{H}^H_k \right\rbrace + \sigma^2_\mathrm{\mathrm{dl}} \mathbf{I}_2    - \mathbb{E}\left\lbrace  \mathbf{H}_k \mathbf{w}_{ki}  \right\rbrace (\mathbb{E}\left\lbrace  \mathbf{H}_k \mathbf{w}_{ki}  \right\rbrace )^H  \right)  \mathbf{v}_{\mathrm{dl},ki}}.
\end{equation}
\hrule

\end{figure*}

\end{lemma}
\begin{IEEEproof}
For polarization $i \in \left\lbrace V, H\right\rbrace $, the downlink signal after receive combining becomes 
\begin{align}
\mathbf{v}_{\mathrm{dl},ki}^H \mathbf{y}_k &= \mathbf{v}^H_{\mathrm{dl},ki} \mathbb{E}\left\lbrace  \mathbf{H}_k \mathbf{w}_{ki}  \right\rbrace  d_{li} \nonumber \\
&+ \mathbf{v}^H_{\mathrm{dl},ki} \left( \mathbf{H}_k \sum_{l=1}^K \mathbf{W}_l \mathbf{d}_l + \mathbf{n}_k -\mathbb{E}\left\lbrace  \mathbf{H}_k \mathbf{w}_{ki}  \right\rbrace  d_{li}  \right).
\end{align}
The first term contains the deterministic factor  $\mathbf{v}^H_{\mathrm{dl},ki} \mathbb{E}\left\lbrace  \mathbf{H}_k \mathbf{w}_{ki}  \right\rbrace$ in front of the desired signal and is uncorrelated with the second term. We can therefore use the capacity lower bound in \cite[Cor.~1.3]{massivemimobook} to obtain the SE expression in \eqref{eq:Rate-woSIC} for polarization $i$.
\end{IEEEproof}

This SE expression can be computed for any choice of precoding and combining vectors. However, the linear MMSE combining vector in \eqref{vMMSE} maximizes it for given precoding vectors.

\subsection{Downlink MMSE-SIC Processing}

Similar to the uplink, the MMSE-SIC scheme can be used to detect signals. The following lemma gives a lower bound on the downlink capacity which makes it an achievable SE.

\begin{lemma}
 An achievable downlink SE of UE $k$ using MMSE-SIC detection  is \cite[Theorem 2]{Li2016b}  
\begin{align} \label{eq:rate-expression}
	\!\!\! \!\! \! R^\mathrm{dl}_k = \frac{\tau_c -\tau_p}{\tau_c}\log_2 \mathrm{det}\left(  \mathbf{I}_2 \!+ \left( \mathbb{E}\left\lbrace  \mathbf{H}_k \mathbf{W}_k  \right\rbrace\right)^H \boldsymbol{\Omega}_k  \mathbb{E}\left\lbrace  \mathbf{H}_k \mathbf{W}_k  \right\rbrace \right), 
\end{align}
where 
\begin{align} 
	 \boldsymbol{\Omega}_k &= 
	 \left(  \mathbb{E}\left\lbrace  \mathbf{H}_k \sum_{l=1}^K \mathbf{W}_l \mathbf{W}^H_l  \mathbf{H}^H_k \right\rbrace    + \sigma^2_\mathrm{dl} \mathbf{I}_2 \right.\nonumber  \\
	 &\left. -\mathbb{E}\left\lbrace  \mathbf{H}_k \mathbf{W}_k  \right\rbrace \left( \mathbb{E}\left\lbrace  \mathbf{H}_k \mathbf{W}_k  \right\rbrace\right)^H   \right)^{-1}.
\end{align}
\end{lemma}
\begin{IEEEproof}
This bound can be achieved if UE $k$ applies  MMSE-SIC detection to $\mathbf{y}_k$ by treating $\mathbb{E}\left\lbrace  \mathbf{H}_k \mathbf{W}_k  \right\rbrace$ as the true channel and the uncorrelated term $\mathbf{y}_k - \mathbb{E}\left\lbrace  \mathbf{H}_k \mathbf{W}_k  \right\rbrace \mathbf{d}_k$ as independent noise.
\end{IEEEproof}
 The rate expression can be computed numerically for any choice of precoding. We consider three linear precoders $\mathbf{W}_k$, namely linear MMSE, zero-forcing and MR that are defined as 

\begin{align}
\mathbf{W}^{X}_k = \begin{bmatrix}
	\frac{\mathbf{v}^{X}_{kV}}{\sqrt{\mathbb{E}\left\lbrace \left\| \mathbf{v}^{X}_{kV}\right\|^2\right\rbrace}}  & \frac{\mathbf{v}^{X}_{kH}}{\sqrt{\mathbb{E}\left\lbrace \left\|\mathbf{v}^{X}_{kH}\right\|^2\right\rbrace}} 
\end{bmatrix}\begin{bmatrix}
	\sqrt{\rho^\mathrm{dl}_{kV} }& 0 \\
	0 &\sqrt{\rho^\mathrm{dl}_{kH} }
\end{bmatrix},
\end{align}
where $X \in \left\lbrace \mathrm{MMSE, ZF, MR}\right\rbrace $.  The corresponding $\mathbf{v}^{X}_{kH}$ vectors are given in \eqref{vMMMSE}-\eqref{vMR}. In the case of MR precoding, the expectations can be computed in closed form as described in the following lemma.

\begin{lemma}\label{lemmaDLMR}
	If MR precoding with 
	\begin{align}
		\mathbf{W}^\mathrm{MR}_k &= \begin{bmatrix}
			\frac{\hat{\mathbf{h}}_{kV}}{\sqrt{\mathbb{E}\left\lbrace \left\| \hat{\mathbf{h}}_{kV}\right\|^2\right\rbrace}}  & \frac{\hat{\mathbf{h}}_{kH}}{\sqrt{\mathbb{E}\left\lbrace \left\| \hat{\mathbf{h}}_{kH}\right\|^2\right\rbrace}} 
		\end{bmatrix}\begin{bmatrix}
			\sqrt{\rho^\mathrm{dl}_{kV} }& 0 \\
			0 &\sqrt{\rho^\mathrm{dl}_{kH} }
		\end{bmatrix} \nonumber \\
		&= \begin{bmatrix}
			\frac{\sqrt{\rho^\mathrm{dl}_{kV} }\hat{\mathbf{h}}_{kV}}{\sqrt{\mathrm{tr}\left(\mathbf{\Gamma}_k^v \right)    }}  & 	\frac{\sqrt{\rho^\mathrm{dl}_{kH} }\hat{\mathbf{h}}_{kH}}{\sqrt{\mathrm{tr}\left(\mathbf{\Gamma}_k^h \right)    }}
		\end{bmatrix}
	\end{align}
	 is used based on the MMSE estimator, then the achievable SE in \eqref{eq:rate-expression} can be computed in closed form as given in \eqref{eqrate}, at the top of next page.
	\begin{figure*}
\begin{align}\label{eqrate}
\!\!\!\!\!\! \!\!\!\!\!	R^\mathrm{dl}_k=\frac{\tau_c -\tau_p}{\tau_c} \left[  \log_2\!\! \left(\! 1 +\frac{	\rho^\mathrm{dl}_{kV} \mathrm{tr}\left(\mathbf{\Gamma}_k^v\right) }{\sum_{l=1}^K \left( \rho^\mathrm{dl}_{lV} \frac{\mathrm{tr}\left( \mathbf{\Gamma}_l^v  \mathbf{R}_{kV}\right)}{\mathrm{tr}\left( \mathbf{\Gamma}_l^v \right)} +\rho^\mathrm{dl}_{lH}\frac{\mathrm{tr}\left( \mathbf{\Gamma}_l^h  \mathbf{R}_{kV}\right)}{\mathrm{tr}\left( \mathbf{\Gamma}_l^h \right)}\right) \! + \! \sigma^2_\mathrm{dl}}\right) 
\!	+ \! \log_2\!\! \left(\! 1 + \frac{	\rho^\mathrm{dl}_{kH} \mathrm{tr}\left(\mathbf{\Gamma}_k^h\right) }{\sum_{l=1}^K \left( \rho^\mathrm{dl}_{lH} \frac{\mathrm{tr}\left( \mathbf{\Gamma}_l^h  \mathbf{R}_{kH}\right)}{\mathrm{tr}\left( \mathbf{\Gamma}_l^h \right)} +\rho^\mathrm{dl}_{lV}\frac{\mathrm{tr}\left( \mathbf{\Gamma}_l^v  \mathbf{R}_{kH}\right)}{\mathrm{tr}\left( \mathbf{\Gamma}_l^v \right)} \right) \! + \! \sigma^2_\mathrm{dl}} \right)\right] . 
\end{align}
\vspace{-1cm}

\end{figure*}

\end{lemma}
\begin{IEEEproof}
The proof follows from direct computation of the expectations and given in Appendix B.
\end{IEEEproof}

The simplified version of \eqref{eqrate} for $ \mathbf{R}_{\mathrm{BS},k} = \beta_k \mathbf{I}_\frac{M}{2}$  for  $k= 1, \dots, K$ is given in \eqref{eqDLsimplified}, at the top of next page.
\begin{figure*}
\begin{align}\label{eqDLsimplified}
	\!\!\!\!\!\!	R^\mathrm{dl}_k&=\frac{\tau_c -\tau_p}{\tau_c}\log_2\left(1 +\frac{\frac{M}{2}\rho_{kV}^\mathrm{dl}(\gamma_{kV,1}   + \gamma_{kV,2} ) }{\displaystyle \sum_{l=1}^K \left(  \rho^\mathrm{dl}_{lV} \frac{\gamma_{lV,1} \beta_k (1 -q_k)   + \gamma_{lV,2} \beta_k q_k}{\gamma_{lV,1}   + \gamma_{lV,2} } + \rho^\mathrm{dl}_{lH} \frac{\gamma_{lV,1} \beta_k q_k  + \gamma_{lV,2} \beta_k (1 -q_k) }{\gamma_{lV,1}  + \gamma_{lV,2}  }\right)   + \sigma_\mathrm{dl}^2  }\right) \nonumber \\
	&+ \frac{\tau_c -\tau_p}{\tau_c} \log_2\left(1 + \frac{\frac{M}{2}\rho_{kH}^\mathrm{dl}(\gamma_{kH,1}   + \gamma_{kH,2} ) }{\displaystyle \sum_{l=1}^K \left(  \rho^\mathrm{dl}_{lH} \frac{\gamma_{lH,1} \beta_k (1 -q_k)   + \gamma_{lH,2} \beta_k q_k}{\gamma_{lH,1}   + \gamma_{lH,2} } + \rho^\mathrm{dl}_{lV} \frac{\gamma_{lH,1} \beta_k q_k  + \gamma_{lH,2} \beta_k (1 -q_k) }{\gamma_{lH,1}  + \gamma_{lH,2}  }\right)   + \sigma_\mathrm{dl}^2  } \right).
\end{align}
\hrule
\end{figure*}

We notice that \eqref{eqrate} and \eqref{eqDLsimplified} have similar structure as \eqref{eqRateUL} and \eqref{eqrateULSimplified}, respectively, in the uplink part. Hence, they can be interpreted similarly.

\section{Power Control}

In this section, we address the problem of maximizing the uplink and downlink sum SEs for MR combining/precoding, based on the new closed-form expressions given in \eqref{eqrateULSimplified} and \eqref{eqDLsimplified}. These expressions were derived for the case of spatially uncorrelated fading, but we will show in Section~\ref{secNumRes} that the obtained solutions work well also in situations with spatial correlation.

\subsection{Uplink Power Control}
\label{ULPC}
First, notice that  the V/H polarizations give equally strong channels, thus it is desirable to make  the uplink pilot powers equal, i.e., $p_l =p_{lV} = p_{lH}$. In this case, the following terms in \eqref{eqrateULSimplified} are symmetric such that 
\begin{align}
&\gamma_{l,1}= \gamma_{lV,1} =  \gamma_{lH,1} =\frac{ p_{l} \tau_p \beta^2_l (1-q_l)^2}{ p_{l} \tau_p \beta_l (1 - q_l ) + \sigma_\mathrm{ul}^2}, \\
&\gamma_{l,2}= \gamma_{lV,2} =  \gamma_{lH,2} =\frac{ p_{l} \tau_p \beta^2_l q_l^2}{ p_{l} \tau_p \beta_l q_l  + \sigma_\mathrm{ul}^2}.
\end{align}
 Motivated by the symmetry of the V/H polarizations,  the same uplink power should also be used at  both polarizations of each UE antenna for data transmission such that  $\rho^\mathrm{ul}_{lV} = \rho^\mathrm{ul}_{lH} =\rho^\mathrm{ul}_{l} $ with $0 \leq\rho^\mathrm{ul}_{l} \leq \rho^\mathrm{ul}_\mathrm{tot}/2$  for $l=1,\dots,K$.    The maximum uplink transmit power at each UE is $\rho^\mathrm{ul}_\mathrm{tot}$. However, it might not be desired to transmit  at maximum power at all UE antennas. Then, we formulate the uplink sum SE maximization problem as   
\begin{equation}\label{ULform1}
\begin{aligned}
	\underset{\rho^\mathrm{ul}_1, \dots, \rho^\mathrm{ul}_K}{\textrm{maximize}}  \quad &  \sum_{k=1}^{K} \frac{2(\tau_c -\tau_p)}{\tau_c}\log_2\left(1 +\frac{	\rho^\mathrm{ul}_k  \frac{M}{2}(\gamma_{k,1}   + \gamma_{k,2} ) }{\displaystyle\sum_{l=1}^K \rho^\mathrm{ul}_l  \beta_l  + \sigma^2_\mathrm{ul}}\right)\\
	\textrm{subject to} \quad & 0 \leq \rho^\mathrm{ul}_l \leq \rho^\mathrm{ul}_\mathrm{tot}/2, \ l=1, \dots, K.\\
\end{aligned}
\end{equation}
 The optimization parameters are $\rho^\mathrm{ul}_1, \dots, \rho^\mathrm{ul}_K$, while  all other terms are constant. The formulation in \eqref{ULform1} is non-convex, but we notice that  the denominator $ \sum_{l=1}^K \rho^\mathrm{ul}_l  \beta_l  + \sigma^2_\mathrm{ul}$ is the same for all UEs. Using this property, we reformulate the problem in convex form similar to  \cite[Theorem 4]{cheng2016optimal} as
\begin{equation}\label{ULform2}
	\begin{aligned}
	\underset{x_1, \dots, x_K, s}{\textrm{maximize}}\quad &  \sum_{k=1}^{K} \frac{2(\tau_c -\tau_p)}{\tau_c}\log_2\left(1 + a^\mathrm{ul}_k x_k\right)\\
		\textrm{subject to} \quad & 0 \leq  x_k \leq  \frac{s \beta_k  \rho^\mathrm{ul}_\mathrm{tot}}{2} , \ k=1, \dots, K\\
		&\sum_{k=1}^K x_k = 1- \sigma_\mathrm{ul}^2 s
	\end{aligned}
\end{equation}
where $
	a^\mathrm{ul}_k =	  \frac{M}{2}\frac{\gamma_{k,1}   + \gamma_{k,2}}{\beta_k}$,
$	s = \frac{1}{ \sum_{l=1}^K \rho^\mathrm{ul}_l  \beta_l  + \sigma_\mathrm{ul}^2}$, 
$ x_k = s \beta_k  \rho^\mathrm{ul}_k $.
Thus, the problems \eqref{ULform1} and  \eqref{ULform2} are equivalent and the solution to \eqref{ULform1} can be obtained from the solution of  \eqref{ULform2} as $ \eta_k = \frac{x_k}{s b^\mathrm{ul}_k}$. Since we have derived the convex problem reformulation \eqref{ULform2}, we can use any general-purpose solver for convex optimization problem to find the optimal solutions efficiently and with guaranteed convergence. In the numerical results section, we use CVX \cite{cvx} and its default solver SDPT3.

\subsection{Downlink Power Control}
\label{DLPC}
Similar to the uplink part, motivated by the symmetry of the  V/H polarizations,  the same downlink power is  used for data transmission at  both polarizations of each UE antenna such that  $\rho^\mathrm{dl}_{lV} = \rho^\mathrm{dl}_{lH} = \rho^\mathrm{dl}_{l} $ with  $ \sum_{l=1}^K \rho^\mathrm{dl}_{l}  \leq \rho^\mathrm{dl}_\mathrm{tot}/2$ and $ \rho^\mathrm{dl}_{l}\geq 0$ for $ l= 1, \dots, K $. The maximum total downlink transmit power at the BS for each polarization is $\rho^\mathrm{dl}_\mathrm{tot}/2$.  Also, the uplink pilot powers are $p_l =p_{lV} = p_{lH}$ as in Section \ref{ULPC}. Then, we formulate the downlink sum SE maximization problem as   
\begin{equation}\label{DLform1}
	\begin{aligned}
		\underset{\rho^\mathrm{dl}_{1} , \dots, \rho^\mathrm{dl}_{K} }{\textrm{maximize}} \quad &  \sum_{k=1}^{K} \frac{2(\tau_c -\tau_p)}{\tau_c}\log_2\left(1 +\frac{	\rho^\mathrm{dl}_k  \frac{M}{2}(\gamma_{k,1}   + \gamma_{k,2} ) }{\beta_k \sum_{l=1}^K  \rho^\mathrm{dl}_l  + \sigma^2_\mathrm{dl}}\right)\\
		\textrm{subject to} \quad & \sum_{l=1}^K\rho^\mathrm{dl}_{l}  \leq  \rho^\mathrm{dl}_\mathrm{tot}/2, \ l=1, \dots, K\\
		&\rho^\mathrm{dl}_{l}\geq 0, \ l=1, \dots, K.\\
	\end{aligned}
\end{equation}
 The optimization parameters are the power allocation coefficients $\rho^\mathrm{dl}_{1}, \dots, \rho^\mathrm{dl}_{K}$ while the other terms are  constant. We notice that the sum rate is larger with $(c\rho^\mathrm{dl}_{1}, \dots,c\rho^\mathrm{dl}_{K})$ than with $(\rho^\mathrm{dl}_{1} , \dots, \rho^\mathrm{dl}_{K})$, for any $c \geq 1$. Hence, the solution to \eqref{DLform1} must use the maximum power $ \sum_{l=1}^K\rho^\mathrm{dl}_l =  \rho^\mathrm{dl}_\mathrm{tot}/2$, and we can rewrite the problem as 
\begin{equation}\label{DLform2}
	\begin{aligned}
		\underset{\eta_1, \dots, \eta_K}{\textrm{maximize}} \quad &  \sum_{k=1}^{K} \frac{2(\tau_c -\tau_p)}{\tau_c}\log_2\left(1 +\frac{	a^\mathrm{dl}_k   }{ b_k^\mathrm{dl} + \sigma^2_\mathrm{dl}} \rho^\mathrm{dl}_k \right)\\
		\textrm{s.t.} \quad & \sum_{k=1}^K \rho^\mathrm{dl}_k   = \rho^\mathrm{dl}_\mathrm{tot}/2, \ k=1, \dots, K\\
		&\rho^\mathrm{dl}_k \geq 0, \ k=1, \dots, K\\
	\end{aligned}
\end{equation}
where $a_k^\mathrm{dl}= 	 \frac{M}{2}(\gamma_{k,1}   + \gamma_{k,2} ) $ and $b_k^\mathrm{dl}= \frac{\rho^\mathrm{dl}_\mathrm{tot} \beta_k}{2}  $. The solutions to this reformulated problem is obtained by the classical water-filling algorithm  as $\rho^\mathrm{dl}_k  =\max\left(\mu - \frac{1 + b_k^\mathrm{dl}}{a_k^\mathrm{dl}}, 0 \right) $, where $\mu$ is the unique solution to $\sum_{k=1}^K\rho^\mathrm{dl}_k   = \rho^\mathrm{dl}_\mathrm{tot}/2$ that is easily found by a line search since the left-hand side is an increasing function of $\mu$ \cite[Ex.~5.2]{Boyd2004a}. This is also the solution to \eqref{DLform1}.

\section{Numerical Results}
 \label{secNumRes}
In this section, we evaluate the performance of dual-polarized antennas under different channel conditions. We consider a single-cell massive MIMO network with $\frac{M}{2}$ dual-polarized antennas and $K=10$ UEs.  The UEs are
independently and uniformly distributed within a square of
size $0.5 \times 0.5$ $\mathrm{km}^2$ at distances larger than 15 m from the BS. The BS is located at the center of the cell. The location of each UE is used when computing the large-scale fading and nominal angle between the UEs and BS.

The BS is equipped with a ULA with half-wavelength
antenna spacing.  For the spatial correlation matrices,
we consider $N_{\textrm{cluster}} = 6$ scattering clusters and the covariance
matrix of each cluster is modeled by the (approximate)
Gaussian local scattering model \cite{ozdogan2019massive} such that
\begin{align} \label{NLoS}
	&\left[ \mathbf{R}_{\mathrm{BS},k}\right] _{s,m} \nonumber \\
	& = \frac{\beta_{k}}{N_{\textrm{cluster}}} \sum_{n=1}^{N_{\textrm{cluster}}} e^{\jmath \pi  (s-m)\sin({\varphi}_{k,n})}  e^{-\frac{\sigma^2_\varphi}{2}\left(\pi (s-m)\cos({\varphi}_{k,n})\right)^2 }, 
\end{align}
where $\beta_{k}$ is the large-scale fading coefficient and ${\varphi}_{k,n} \sim \mathcal{U}[{\varphi}_{k}-40^\circ, \ {\varphi}_{k} + 40^\circ]$ is the nominal angle of arrival (AoA) for the $n$ cluster. The multipath components of a cluster have Gaussian distributed AoAs, distributed around the nominal AoA with the angular standard deviation (ASD) $\sigma_\varphi=5^\circ$. 
Note that \eqref{NLoS} is an approximate closed-form expression of a more general integral expression from \cite[Ch.~2]{massivemimobook} and the approximation is accurate for $\sigma_\varphi<15^\circ$, which is satisfied here.
The large-scale fading coefficient is modeled (in dB) as 
\begin{equation}\label{betak}
	\beta_{k}= -35.3 -37.6  \log_{10}\left( \frac{{d_{k}}}{\mathrm{1 m}} \right) + F_{k},
\end{equation}
where $d_k$ is the distance between the BS and UE $k$, $F_{k} \sim \mathcal{N}(  0, \sigma^2_\mathrm{sf} )$ is the shadow fading with $\sigma_\mathrm{sf} = 7$.  

We consider communication over a $20$ MHz channel and the total receiver noise power is $-94$ dBm. Each coherence block consists of $\tau_c$ = $200$ samples and $\tau_p = 2K = 20$ pilots are allocated for  channel estimation. Unless otherwise stated, equal power allocation is applied such that the pilot powers are $p_{kV}= p_{kH} = 100$ mW and the uplink and downlink transmit powers are $\rho^\mathrm{ul}_{kV}= \rho^\mathrm{ul}_{kH} =\rho^\mathrm{dl}_{kV}= \rho^\mathrm{dl}_{kH} = 100$ mW for every UE $k =1,\dots, K$. Also, $\rho^\mathrm{ul}_\mathrm{tot} = 200$ mW and $\rho^\mathrm{dl}_\mathrm{tot} = 2K \times 100$ mW. The XPD value is 5 dB for all UEs.

\begin{figure}[h!]
	\centering
	\includegraphics[scale=0.5]{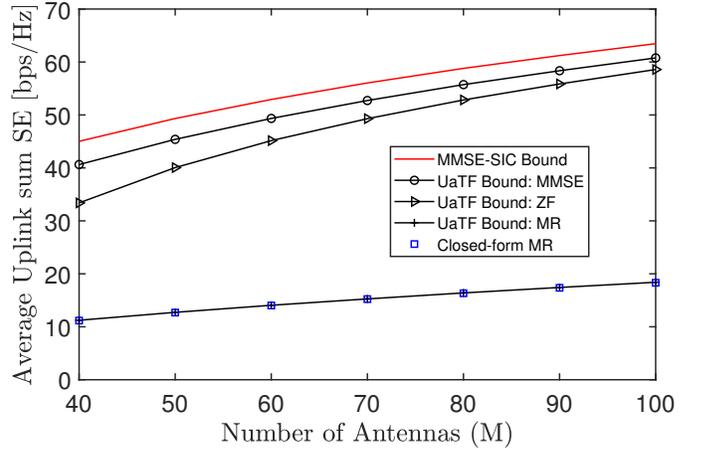} 
	\caption{Average uplink sum SE for $10$ UEs with different combining schemes.}\label{figUplink}
\end{figure}

\begin{figure}[h!]	
	\centering
	\includegraphics[scale=0.5]{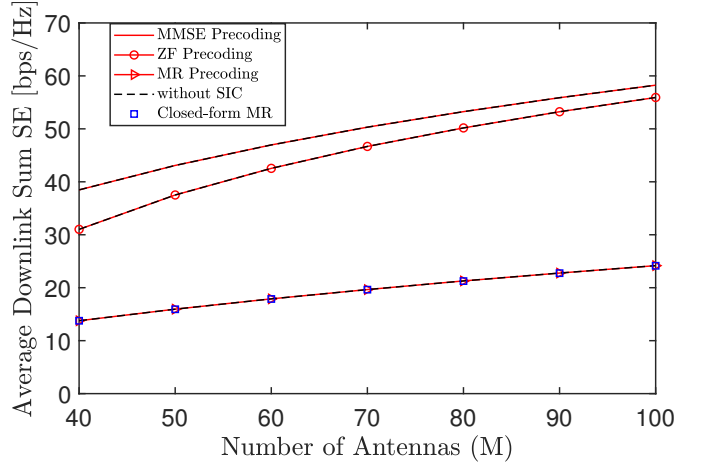} 
	\caption{Average downlink sum SE for $10$ UEs with different precoding schemes.}\label{figDownlink}
\end{figure}

\emph{1) Performance Comparison of Different Combining/Precoding Schemes:} In Fig.~\ref{figUplink}, the uplink sum SE of MMSE-SIC and UaTF bound with MMSE, ZF and MR detectors are shown. The SEs are averaged over different UE locations and shadow fading realizations. The highest SE is achieved by using MMSE-SIC scheme, as expected. Yet, the linear MMSE scheme achieves a competitive uplink sum SE.  It shows that a linear detector can reach most of the SE  from employing a dual-polarized antenna at the UE side.  We also observe that the performance achieved with the MMSE and ZF combining vectors are significantly better than MR. It is because of the fact that the MR combining vector does not have the ability to cancel  inter-stream interference. Fig.~\ref{figDownlink} shows the downlink sum SE for the MMSE, ZF and MR precoders. The solid lines denote the cases where the UEs apply MMSE-SIC to the received data streams, whereas the dashed lines are for the linear MMSE combining. At each UE, MMSE-SIC detection is applied to the two data streams that are received by the orthogonally polarized antennas. The MMSE-SIC and linear MMSE schemes give the  same performance due to the lack of polarization correlation between the received data streams. 
The SE grows monotonically with the number of antennas when using any of the processing schemes, which indicates that one can reach any desired SE value by deploying sufficiently many BS antennas; however, fewer antennas are needed when using more advanced combining/precoding schemes.


\begin{figure}[h!]
	\centering
	\includegraphics[scale=0.5]{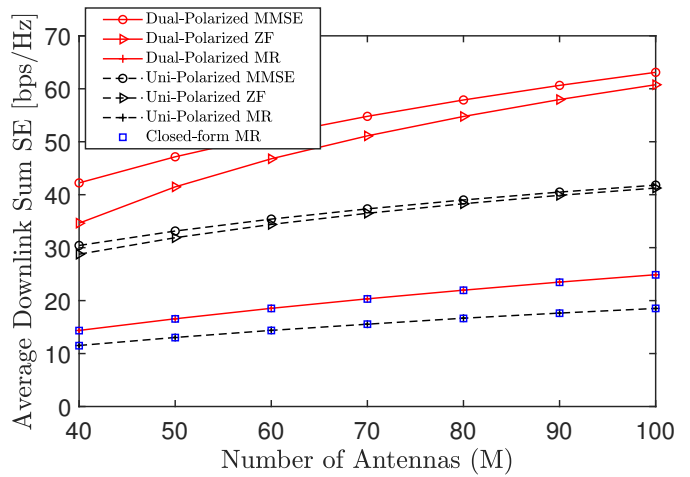} 
	\caption{Average downlink sum SE for 10 UEs with different precoders as a function of the number of BS antennas for dual-polarized and uni-polarized setups with $M_\mathrm{uni} = M$ and $p_\mathrm{uni} = \rho^\mathrm{dl}_\mathrm{uni} = 100$ mW.  }\label{figDPvsUP}
\end{figure}

\begin{figure}[h!]
	\centering
	\includegraphics[scale=0.5]{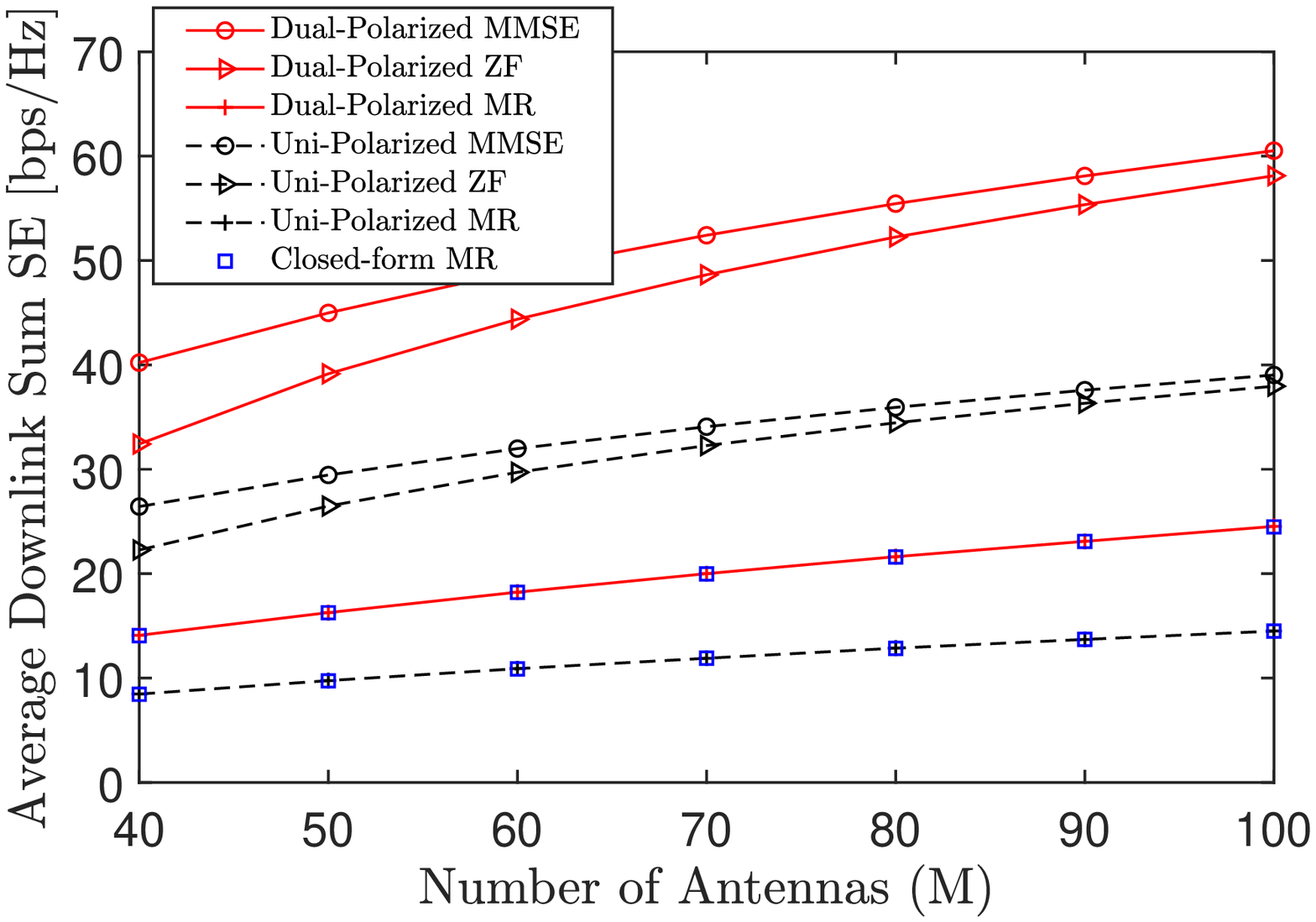} 
	\caption{Average downlink sum SE for 10 UEs with different precoders as a function of the number of BS antennas for dual-polarized and uni-polarized setups with $M_\mathrm{uni} = \frac{M}{2}$ and $p_\mathrm{uni} = \rho^\mathrm{dl}_\mathrm{uni} = 200$ mW.  }\label{figDPvsUP2}
\end{figure}

\emph{2) Dual-Polarized vs Uni-Polarized Antennas:} Fig.~\ref{figDPvsUP} and Fig.~\ref{figDPvsUP2} compare the sum downlink SEs with dual-polarized and uni-polarized antennas. In the uni-polarized benchmark, we consider a single antenna per UE and denote the number of BS antennas as $M_\mathrm{uni}$. In the case of $M_\mathrm{uni}=\frac{M}{2}$,  the H polarized antennas are removed from  the dual-polarized antenna arrays both at the BS and UE sides. Therefore, the number of antennas is halved whereas the total array aperture remains the same in both setups. In contrast, $M_\mathrm{uni}= M$, the array is made of only V polarized antennas. Thus, the total array aperture is doubled compared to the dual-polarized setup.

We consider MMSE, ZF and MR precoding  for both antenna setups. For the dual-polarized case,  MMSE-SIC decoding at the UE side is considered. In the uni-polarized antenna setup, we implemented the MMSE, ZF and MR precoding vectors as

\begin{equation}
\!\!\!\!\!\!\!\!\!\!\!\!\!\!\!	\mathbf{W}^\mathrm{uni,MMSE}_k \!\!\!\!\!\!\!\!\!= 
	\frac{\sqrt{\rho^\mathrm{dl}_\mathrm{uni}} \left( \sum_{l = 1}^K \rho^\mathrm{ul}_\mathrm{uni} \hat{\mathbf{h}}_l \hat{\mathbf{h}}_l^H + \rho^\mathrm{ul}_\mathrm{uni}\mathbf{C}_l + \sigma_\mathrm{ul}^2 \mathbf{I}_{M_\mathrm{uni}}\!\right)^{-1} \!\! \hat{\mathbf{h}}_k}{\!\!\sqrt{\!\mathbb{E}\left\lbrace \left\|  \left( \sum_{l = 1}^K \rho^\mathrm{ul}_\mathrm{uni} \hat{\mathbf{h}}_l \hat{\mathbf{h}}_l^H + \rho^\mathrm{ul}_\mathrm{uni}\mathbf{C}_l + \sigma_\mathrm{ul}^2 \mathbf{I}_{M_\mathrm{uni}}\!\right)^{-1}\!\! \hat{\mathbf{h}}_k\right\|^2\!\right\rbrace}},
\end{equation}
\begin{equation}
	\mathbf{W}^\mathrm{uni, ZF}_k = 
	\frac{\sqrt{\rho^\mathrm{dl}_\mathrm{uni}}[\mathbf{W}_\mathrm{uni,all}]_{k}}{ \sqrt{\mathbb{E}\left\lbrace\left\|  [\mathbf{W}_\mathrm{uni, all}]_{k}\right\|^2\right\rbrace} }, 
\end{equation}
\begin{equation}
	\mathbf{W}^\mathrm{uni}_k = 
	\frac{\sqrt{\rho^\mathrm{dl}_\mathrm{uni}} \hat{\mathbf{h}}_k}{\sqrt{\mathbb{E}\left\lbrace \left\| \hat{\mathbf{h}}_{k}\right\|^2\right\rbrace}},
\end{equation}
where $\mathbf{W}_\mathrm{uni, all}= \mathbf{H}_\mathrm{uni,all} \left( \mathbf{H}^H_\mathrm{uni,all}\mathbf{H}_\mathrm{uni,all}\right)^{-1} $,   $\mathbf{H}_\mathrm{uni,all}= [\hat{\mathbf{h}}_1, \dots, \hat{\mathbf{h}}_k, \dots  \hat{\mathbf{h}}_K  ] \in \mathbb{C}^{M_\mathrm{{uni}} \times K}$. The precoding vectors are generated based on the MMSE estimates of channels ${\mathbf{h}}_k \sim \mathcal{N}_\mathbb{C}(  \mathbf{0}, \mathbf{R}_{\mathrm{BS},k}  )$ as  $\hat{\mathbf{h}}_k \sim \mathcal{N}_\mathbb{C}(  \mathbf{0},p_\mathrm{uni} \tau_\mathrm{uni,p}\mathbf{R}_{\mathrm{BS},k}  \mathbf{\Psi}_k\mathbf{R}_{\mathrm{BS},k} )$ with $\mathbf{\Psi}_k = \left( p_\mathrm{uni} \tau_\mathrm{uni,p}\mathbf{R}_{\mathrm{BS},k} + \sigma^2_\mathrm{ul} \mathbf{I}_{M_\mathrm{uni}} \right)^{-1}  $ and $\mathbf{C}_k$ is the estimation error covariance matrix, see \cite[Sec. 4]{massivemimobook} for the details. To have a fair comparison, we set $p_\mathrm{uni} = \rho^\mathrm{ul}_\mathrm{uni}=\rho^\mathrm{dl}_\mathrm{uni} = 200$ mW if $M_\mathrm{uni} = \frac{M}{2}$ and $p_\mathrm{uni} = \rho^\mathrm{ul}_\mathrm{uni}=\rho^\mathrm{dl}_\mathrm{uni} = 100$ mW if $M_\mathrm{uni} = M$.  Besides, $\tau_\mathrm{uni,p} = 10$ so that the total power is constant and the pilot lengths are minimized in both setups.
The SE expressions from \cite[Sec.~4.3]{massivemimobook} are utilized to calculate the downlink SE with MR precoding for uni-polarized antennas. 

By utilizing the two polarization dimensions, the dual-polarized systems can ideally double the multiplexing gain, as the signal-to-noise ratio (SNR) goes to infinity. In Fig.~\ref{figDPvsUP}, the downlink sum SE with dual-polarized and uni-polarized antenna arrays are depicted where $M_\mathrm{uni} = M$ and $p_\mathrm{uni} = \rho_\mathrm{uni} = 100$ mW. We observe that the dual-polarized setup offers better performance than the uni-polarized setup.  The ratios between the average sum SEs of the dual-polarized and uni-polarized setups are approximately $1.5$ for MMSE, $1.4$ for ZF precoding and $1.3$ for MR precoding. In Fig.~\ref{figDPvsUP2} where $M_\mathrm{uni} = M/2$, it is seen that the ratios between the average sum SEs of the dual-polarized and uni-polarized setups are approximately $1.6$ for MMSE and ZF precoding and $1.7$ MR precoding. Note that the ratio is not equal to 2 because the XPD is finite (meaning that there is a polarization leakage), the SNR is finite, and the prelog factors  $(\tau_c- \tau_p )/\tau_c$ and $(\tau_c- \tau_\mathrm{uni,p} )/\tau_c$ are different since half the numbers of pilots are used to estimate the uni-polarized channels. The fact that the markers overlap with the curves confirms the validity of our analytical results in Lemma \ref{lemmaDLMR}. The same behaviors are observed in the uplink but are omitted to avoid repetition.

\begin{figure}[h!]
	\centering
	\includegraphics[scale=0.5]{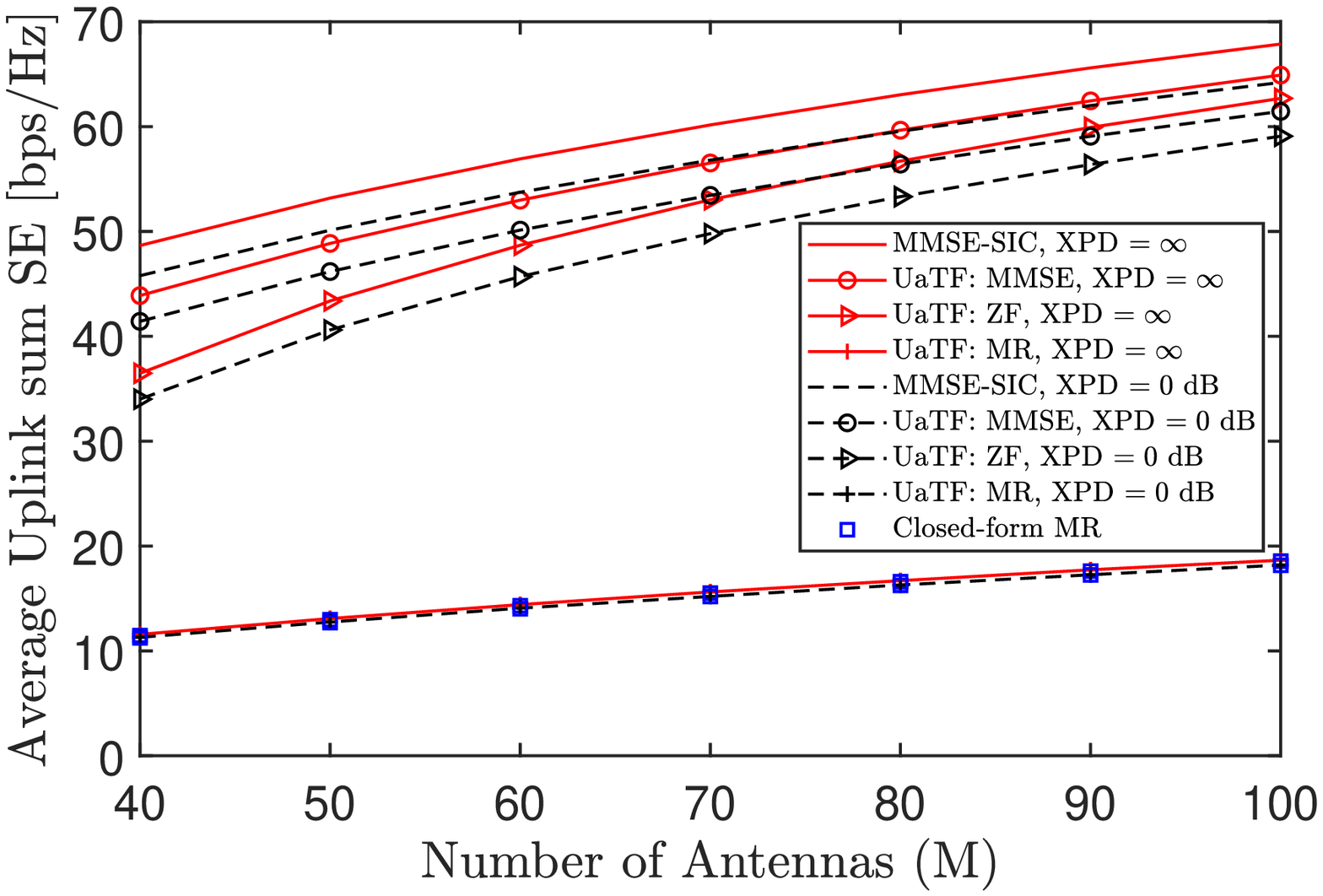} 
	\caption{Average uplink sum SE for 10 UEs with different precoders as a function of the number of BS antennas for different XPD values.} \label{fig3}
\end{figure}

\begin{figure}[h!]
	\centering
	\includegraphics[scale=0.5]{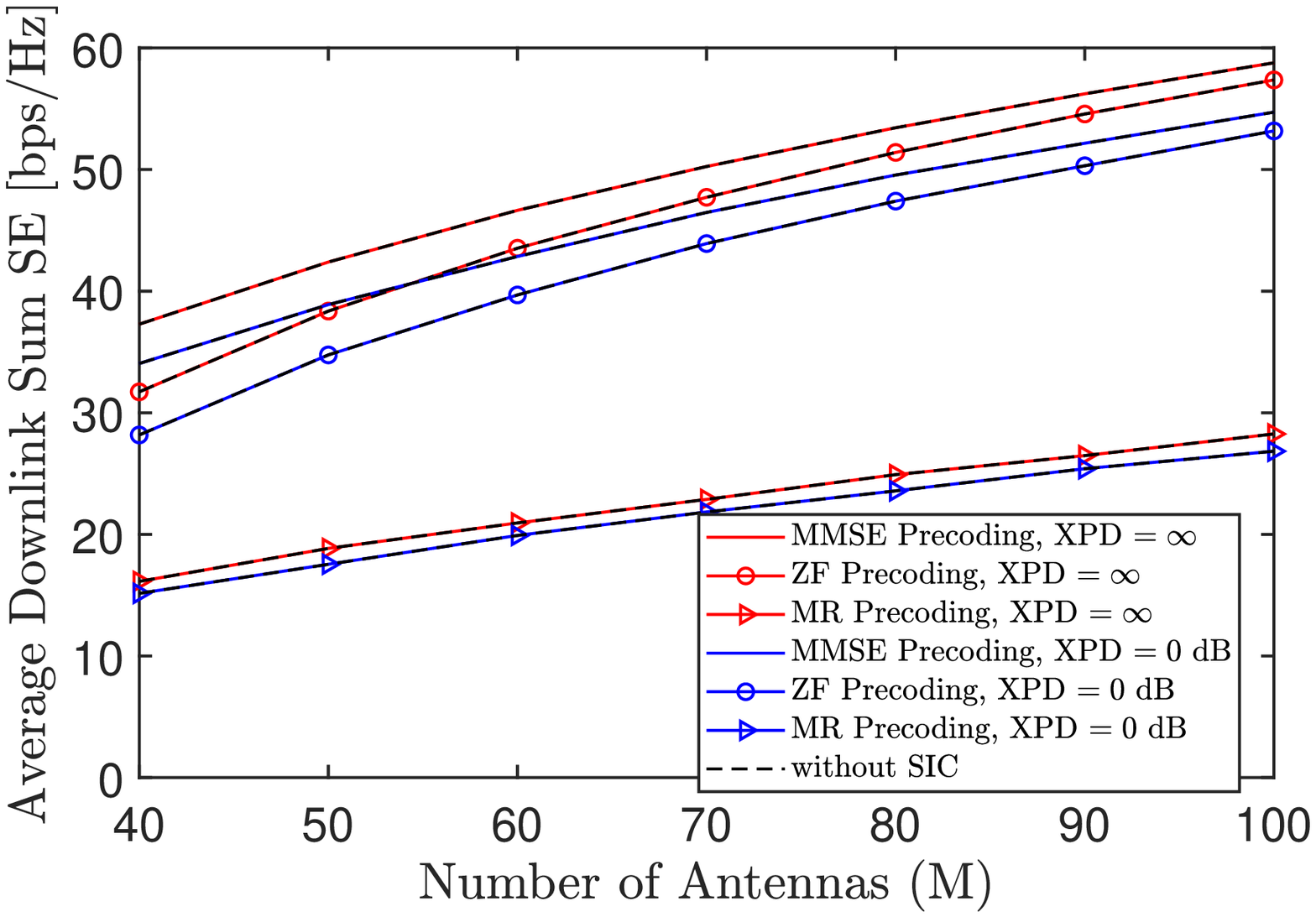} 
	\caption{Average downlink sum SE for 10 UEs with different precoders as a function of the number of BS antennas for different XPD values.} \label{fig4}
\end{figure}

\begin{figure}[h!]
	\centering
	\includegraphics[scale=0.5]{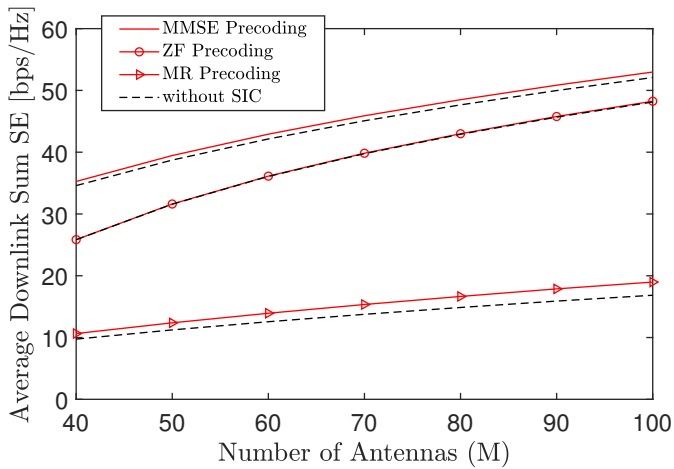} 
	\caption{Average downlink sum SE in the same setup as Fig.~\ref{figDownlink}, except that the XPC coefficients are $t_{k} = r_{k}= 0.8$ for all $k$.}\label{figDownlinkXPC}
\end{figure}

 \emph{3) Effect of Channel Polarization Leakage (XPD):} Fig.~\ref{fig3} shows the average sum uplink SE for two extreme cases of XPD: $\mathrm{XPD}_k=1=0$\,dB (half-power leakage) and $\mathrm{XPD}_k = \infty$ (no leakage). The same XPD values are used across UEs. We observe that the SEs are higher when there is no leakage but the SE difference is only 5\%, thus we can conclude the existence of XPD will have a limited impact on the SE gains achievable using dual-polarized antennas. We further observe that the gap between  MMSE-SIC and linear MMSE increases with the polarization leakage. In Fig.~\ref{fig4}, we see the effect of channel XPD on the downlink SEs. Similar to the uplink, the downlink SEs are higher when there is no leakage (XPD is infinite) and the SE loss is now 5-10\%. One reason for the small SE loss is that the receiver processing can partially compensate for the XPD by using the received signals from both polarizations when decoding the signal that was meant to only be transmitted over one of the polarization dimensions. This feature is not available in systems with uni-polarized antennas, where the power leakage due to XPD is lost.

\emph{4) Effect of Polarization Correlations (XPC):} Fig.~\ref{figDownlinkXPC} shows the average sum downlink SE when the XPC terms at both the transmitter and receiver side are set to $t_k =r_k =0.8$ for $k=1,\ldots,K$. Recall that these variables were taken as $t_k =r_k =0$ in the previous plots. It is seen that there is now a gap between the cases with and without SIC, in contrast to Fig.~\ref{figDownlink} where the same setup with no polarization correlation is present. Also, compared to Fig.~\ref{figDownlink}, we observe that a high correlation between the polarized waves reduces the average sum SEs by 15-25\% (but when XPC exists, it is substantially smaller than $0.8$ \cite[Table 1]{oestges2008dual}). The ZF precoder  is the least effected from a nonzero XPC among the precoders.

\begin{figure}[h!]	
	\centering
	\includegraphics[scale=0.5]{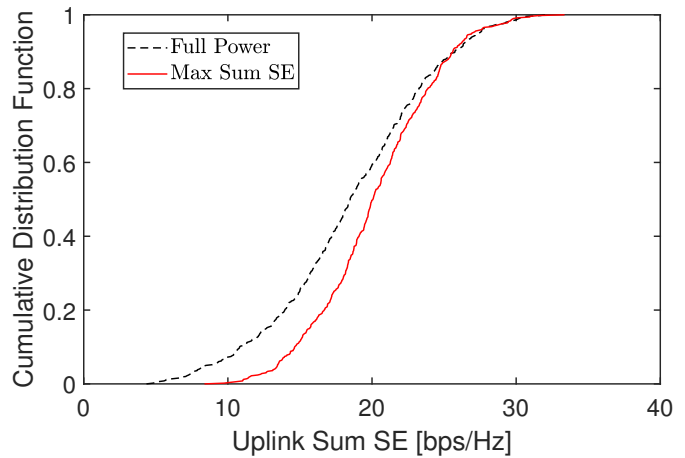} 
	\caption{Cumulative distribution function of uplink sum SE for MR combining scheme with $M=100$.}\label{figUplinkPC}
\end{figure}

\begin{figure}[h!]	
	\centering
	\includegraphics[scale=0.5]{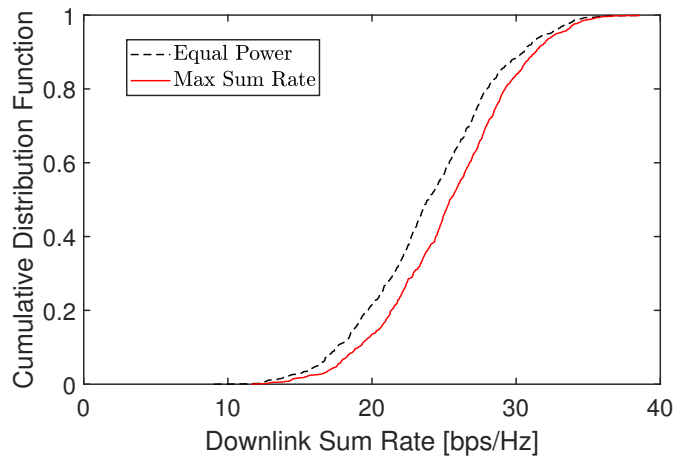} 
	\caption{Cumulative distribution function of downlink sum SE for MR precoding scheme with $M=100$.}\label{figDownlinkPC}
\end{figure}

\emph{4) Power Control for Uplink and Downlink SE with MR:}
In Fig.~\ref{figUplinkPC} and Fig.~\ref{figDownlinkPC}, we compare the sum SE achieved by the benchmark equal power allocation (Full Power/Equal Power) method with the uplink and downlink power control schemes for MR combining/precoding (Max Sum SE) that are described in Section \ref{ULPC} and \ref{DLPC}, respectively. Note that the algorithms are designed by assuming uncorrelated fading, but we apply them in a scenario with correlated fading. In the uplink (Fig.~\ref{figUplinkPC}), we observe that the Max Sum SE scheme increases the sum SE compared to the full power scheme. It shows that when some of the UEs cut down their transmit power, it helps to mitigate the interference that they are creating and improves the uplink sum SE. In the downlink (Fig.~\ref{figDownlinkPC}),  Max Sum SE provides an improvement over the equal power scheme by allocating more power to the better channels and less power to the weaker channels. This improves the sum SE for all UE realizations.

\section{Generalization to Multiple Dual-Polarized UE Antennas}

The previous sections considered the case where each UE is equipped with a single dual-polarized antenna. There are two main reasons for this assumption.
Firstly, to establish baseline SE formulas that are analytically tractable for resource allocation optimization. Secondly, multi-user MIMO in 5G only support two streams per UE, even if the device has more antennas. In uni-polarized multi-user scenarios where each UE has multiple antennas, it is preferred to use the extra antennas to improve the channel conditions rather than spatial multiplexing \cite{Bjornson2013b}. The natural dual-polarized extension is to transmit one stream per polarization and use the extra antennas to improve the channel conditions.
In this section, we will describe how to manage that case within the scope of this paper. 

If UE $k$ is equipped with $N/2$ dual-polarized antennas, its channel matrix from \eqref{eq6} can be generalized to $\mathbf{H}_k \in \mathbb{C}^{N \times M}$ and expressed as 
\begin{align}\label{eq6-multiantenna-user}
	\mathbf{H}_k  = \begin{bmatrix}
	 \mathbf{R}_{\mathrm{UE},k}^{1/2}\mathbf{S}_{kV} \mathbf{R}_{kV}^{1/2} \\
		 \mathbf{R}_{\mathrm{UE},k}^{1/2} \mathbf{S}_{kH}  \mathbf{R}_{kH}^{1/2}
	\end{bmatrix}
	\end{align}
	where $\mathbf{R}_{\mathrm{UE},k} \in \mathbb{C}^{\frac{N}{2} \times \frac{N}{2}}$ is the spatial correlation matrix at the UE side and 
	$\mathbf{S}_{kV},\mathbf{S}_{kV}  \in \mathbb{C}^{\frac{N}{2} \times M}$ have i.i.d.~$\mathcal{N}_\mathbb{C}(0, 1 )$-entries.
	This relatively compact channel matrix notation captures both the spatial channel correlation and the XPD coefficients, thanks to the covariance matrices derived in \eqref{eq:Rvk}-\eqref{eq5} and by letting the first $N/2$ rows in \eqref{eq6-multiantenna-user} represent the V polarized UE antenna elements and the last $N/2$ rows represent the H polarized UE antenna elements.

If the UE will only transmit one uplink stream per polarization, it can transmit them over its multiple antennas using a fixed precoder matrix $\mathbf{M}_k \in \mathbb{C}^{\frac{N}{2} \times 2}$. Suppose $\lambda_{k}$ is the dominant eigenvalue of $\mathbf{R}_{\mathrm{UE},k}$ and that $\mathbf{u}_k  \in \mathbb{C}^{\frac{N}{2} \times 1}$ is the corresponding normalized eigenvector. The UE can utilize this statistical information to perform eigenbeamforming with
\begin{equation}
\mathbf{M}_k = \begin{bmatrix} \mathbf{u}_k & \mathbf{0} \\ \mathbf{0} & \mathbf{u}_k \end{bmatrix}.
\end{equation}
This will create an uplink MIMO channel with the effective channel matrix
\begin{align}
	\mathbf{M}_k^H \mathbf{H}_k  = \begin{bmatrix}
	  \mathbf{u}_k^H \mathbf{R}_{\mathrm{UE},k}^{1/2}\mathbf{S}_{kV} \mathbf{R}_{kV}^{1/2} \\
		 \mathbf{u}_k^H \mathbf{R}_{\mathrm{UE},k}^{1/2} \mathbf{S}_{kH}  \mathbf{R}_{kH}^{1/2}
	\end{bmatrix}  = \lambda_{k}^{1/2} \begin{bmatrix}
	   \mathbf{s}_{kV}^H \mathbf{R}_{kV}^{1/2} \\
		 \mathbf{s}_{kH}^H  \mathbf{R}_{kH}^{1/2}
	\end{bmatrix} 
	\end{align}
where we used the notation 
$\mathbf{s}_{kV} = \mathbf{S}_{kV}^H \mathbf{u}_k$ and $\mathbf{s}_{kH} = \mathbf{S}_{kH}^H \mathbf{u}_k$ and notice that these vectors have  i.i.d.~$\mathcal{N}_\mathbb{C}(0, 1 )$-entries.
Interestingly, this is a channel matrix of the same type as considered previously in the paper, except for the extra factor $\lambda_{k}^{1/2}$ that represents the uplink beamforming gain. This gain is obtained without the need for changing the channel estimation procedures.
Hence, the theory developed in previous chapters can be directly applied to characterize the uplink SEs, if one just scales the channel matrix correctly. If eigenbeamforming is used for reception in the downlink, then the same downlink SE expressions can also be achieved, except for the extra scaling factor $\lambda_{k}^{1/2}$.

\section{Conclusions}

This paper studied  a single-cell massive MIMO system with dual-polarized antennas at both the BS and UEs.  We analyzed uplink and downlink achievable SEs with and without SIC for the linear MMSE, ZF and MR combining/precoding schemes. It is observed that the MMSE-SIC scheme gives a better performance in the uplink whereas linear precoding performs the same as MMSE-SIC in the downlink. In addition, we derived the MMSE channel estimator and characterized its statistics. Using the estimates for MR combining/precoding, we computed  closed-form uplink and downlink SEs. The SE expressions provide insights into the operation and interference behavior when having dual-polarized channels.  Besides, uplink and downlink power control algorithms based on these closed-form expressions are developed.

The dual-polarized and uni-polarized antenna setups are compared numerically. Moreover, the impact of XPD and XPC  on uplink and downlink SEs are evaluated. The expression shows how the multiplexing gain can be doubled by utilizing the polarization domain. We observe that dual-polarized arrays have the same physical size and beamforming gain per polarization as a uni-polarized array with half the number of antennas. Hence, the size can be reduced while maintaining or improving the SE.

\section{Appendix}

\subsection{Proof of Lemma \ref{lemmaULSIC}}
Assume that the signals are decoded in an arbitrary order $x_{1V}, x_{1H}, \dots, x_{KV}, x_{KH}$. First, we can rewrite \eqref{eqULreceived} as
\begin{align}\label{eqULrec}
	\mathbf{y} =& \sum_{l=1 }^K \sqrt{p_{lV}} \hat{\mathbf{h}}_{lV} x_{lV} + \sqrt{p_{lH}} \hat{\mathbf{h}}_{lH} x_{lH}  + \tilde{\mathbf{H}}^H_l \mathbf{P}^{1/2}_l  \mathbf{x}_l   + \mathbf{n} \nonumber \\
	=&\sqrt{p_{1V}} \hat{\mathbf{h}}_{1V} x_{1V} + \mathbf{n}_{1V},
\end{align}
where the first signal ${x}_{1V} $ is transmitted through the effective channel $\sqrt{p_{1V}} \hat{\mathbf{h}}_{1V} $ that is known at the BS and 
$\mathbf{n}_{1V} = \mathbf{y} -  \sqrt{p_{1V}} \hat{\mathbf{h}}_{1V} x_{1V} =  \sum_{l = 2 }^K \sqrt{p_{lV}} \hat{\mathbf{h}}_{lV} x_{lV} + \sum_{l=1 }^K \sqrt{p_{lH}} \hat{\mathbf{h}}_{lH} x_{lH}  + \tilde{\mathbf{H}}^H_l \mathbf{P}^{1/2}_l  \mathbf{x}_l   + \mathbf{n}  $ is the uncorrelated colored noise and interference. The  noise term $\mathbf{n}_{1V} $ has zero mean and its conditional covariance matrix is
\begin{align}
	\mathbf{\Upsilon}_{1V} =&\mathbb{E}\left\lbrace \mathbf{n}_{1V} \mathbf{n}^H_{1V} | \hat{\mathbf{H}}_1,  \dots, \hat{\mathbf{H}}_K \right\rbrace\nonumber \\
	 =&  \sum_{l = 2}^K   {p_{lV}} \hat{\mathbf{h}}_{lV} \hat{\mathbf{h}}^H_{lV} + \sum_{l = 1}^K   {p_{lH}} \hat{\mathbf{h}}_{lH} \hat{\mathbf{h}}^H_{lH}  +p_{lV} \mathbf{C}^v_l \nonumber \\
	 &+ p_{lH} \mathbf{C}^h_l + \sigma^2_\mathrm{ul} \mathbf{I}_{M}.
\end{align}
The first stream $x_{1V}$ is detected using MMSE filter $ \mathbf{\Upsilon}^{-1}_{1V}\hat{\mathbf{h}}_{1V}$ where the corresponding instantaneous SNR is $\mathrm{SNR}_{1V}=p_{1V}\hat{\mathbf{h}}^H_{1V}\mathbf{\Upsilon}_{1V}^{-1}\hat{\mathbf{h}}_{1V}$. After decoding  $x_{1V}$,  it is removed from the received signal $\mathbf{y}$.  Thus, the next data stream $x_{1H}$ is detected based on
\begin{align}
	\mathbf{y} -  \sqrt{p_{1V}} \hat{\mathbf{h}}_{1V} x_{1V} =& \sqrt{p_{1H}} \hat{\mathbf{h}}_{1H} x_{1H}  +\sum_{l = 2 }^K \sqrt{p_{lV}} \hat{\mathbf{h}}_{lV} x_{lV} \nonumber \\
	+& \sqrt{p_{lH}} \hat{\mathbf{h}}_{lH} x_{lH}  
	+  \sum_{l=1 }^K \tilde{\mathbf{H}}^H_l \mathbf{P}^{1/2}_l  \mathbf{x}_l   + \mathbf{n} \nonumber \\
	&= \sqrt{p_{1H}} \hat{\mathbf{h}}_{1H} x_{1H}  + \mathbf{n}_{1H}
\end{align}
using the MMSE filter $ \mathbf{\Upsilon}_{1H}^{-1}\hat{\mathbf{h}}_{1H}$ where the corresponding instantaneous SNR is $\mathrm{SNR}_{1H} =p_{1H}\hat{\mathbf{h}}^H_{1H} \mathbf{\Upsilon}_{1H}^{-1}\hat{\mathbf{h}}_{1H}$ with 
\begin{align}
	\mathbf{\Upsilon}_{1H} = & \mathbb{E}\left\lbrace \mathbf{n}_{1H} \mathbf{n}^H_{1H} | \hat{\mathbf{H}}_1,  \dots, \hat{\mathbf{H}}_K \right\rbrace \nonumber \\
	=&  \sum_{l = 2}^K   {p_{lV}} \hat{\mathbf{h}}_{lV} \hat{\mathbf{h}}^H_{lV} + {p_{lH}} \hat{\mathbf{h}}_{lH} \hat{\mathbf{h}}^H_{lH}+ \sum_{l = 1}^K     p_{lV} \mathbf{C}^v_l \nonumber \\
	 &+p_{lH} \mathbf{C}^h_l + \sigma^2_\mathrm{ul} \mathbf{I}_{M}.
\end{align}
This process is repeated for all $2K$ signals. The last two signals $x_{KV}$ and $x_{KH}$ are decoded as 
\begin{align}
	\mathbf{y} -&  \sum_{l=1}^{K-1}\left( \sqrt{p_{lV}} \hat{\mathbf{h}}_{lV} x_{lV} + \sqrt{p_{lH}} \hat{\mathbf{h}}_{lH} x_{lH} \right)  =  \sqrt{p_{KV}} \hat{\mathbf{h}}_{KV} x_{KV} \nonumber \\
	&+ \sqrt{p_{KH}} \hat{\mathbf{h}}_{KH} x_{KH}  +  \sum_{l=1 }^K \tilde{\mathbf{H}}^H_l \mathbf{P}^{1/2}_l  \mathbf{x}_l   + \mathbf{n} \nonumber \\
	&= \sqrt{p_{KV}} \hat{\mathbf{h}}_{KV} x_{KV}  + \mathbf{n}_{KV}
\end{align}
with 
\begin{align}
	\mathbf{\Upsilon}_{KV} = & \mathbb{E}\left\lbrace \mathbf{n}_{KV} \mathbf{n}^H_{KV} | \hat{\mathbf{H}}_1,  \dots, \hat{\mathbf{H}}_K \right\rbrace \nonumber \\
	= &   {p_{KH}} \hat{\mathbf{h}}_{KH} \hat{\mathbf{h}}^H_{KH}+ \sum_{l = 1}^K     p_{lV} \mathbf{C}^v_l +p_{lH} \mathbf{C}^h_l + \sigma^2_\mathrm{ul} \mathbf{I}_{M},
\end{align}
and
\begin{align}
	\mathbf{y} & -  \sum_{l=1}^{K} \sqrt{p_{lV}} \hat{\mathbf{h}}_{lV} x_{lV} +  \sum_{l=1}^{K-1}\sqrt{p_{lH}} \hat{\mathbf{h}}_{lH} x_{lH}\nonumber\\
	   &=  \sqrt{p_{KH}} \hat{\mathbf{h}}_{KH} x_{KH}  +  \sum_{l=1 }^K \tilde{\mathbf{H}}^H_l \mathbf{P}^{1/2}_l  \mathbf{x}_l   + \mathbf{n} \nonumber \\
	&= \sqrt{p_{KV}} \hat{\mathbf{h}}_{KV} x_{KV}  + \mathbf{n}_{KH},
\end{align}
with  
\begin{align}
	\mathbf{\Upsilon}_{KH} =\mathbb{E}\left\lbrace \mathbf{n}_{KH} \mathbf{n}^H_{KH} \right\rbrace =   \sum_{l = 1}^K     p_{lV} \mathbf{C}^v_l +p_{lH} \mathbf{C}^h_l + \sigma^2_\mathrm{ul} \mathbf{I}_{M}.
\end{align}

Then, the achievable SE of UE $k$ becomes  \cite[Chapter 8]{Tse2005a}
\begin{align}\label{eqRkSIC}
		R^\mathrm{ul,SIC}_k \!\!\!
	&= \frac{\tau_c -\tau_p}{\tau_c}\mathbb{E} \left\lbrace \log_2 (1 + p_{kV}\hat{\mathbf{h}}^H_{kV} \mathbf{\Upsilon}_{kV}^{-1}\hat{\mathbf{h}}_{kV}) \right.\nonumber \\
	&+ \left. \log_2(1 + p_{kH}\hat{\mathbf{h}}^H_{kH} \mathbf{\Upsilon}_{kH}^{-1}\hat{\mathbf{h}}_{kH} ) \right\rbrace \nonumber \\
	\stackrel{(a)}{=}&\frac{\tau_c -\tau_p}{\tau_c}\mathbb{E}\left\lbrace \log_2 \mathrm{det} ( p_{kV}\hat{\mathbf{h}}_{kV}\hat{\mathbf{h}}^H_{kV} +\mathbf{\Upsilon}_{kV})- \log_2 \mathrm{det} (\mathbf{\Upsilon}_{kV}) 	\right\rbrace \nonumber\\
	&+ \frac{\tau_c -\tau_p}{\tau_c}\mathbb{E}\left\lbrace \log_2 \mathrm{det} ( p_{kH}\hat{\mathbf{h}}_{kH}\hat{\mathbf{h}}^H_{kH} +\mathbf{\Upsilon}_{kH})- \log_2 \mathrm{det} (\mathbf{\Upsilon}_{kH}) 	\right\rbrace \nonumber\\
	\stackrel{(b)}{=}&\frac{\tau_c -\tau_p}{\tau_c}\mathbb{E}\left\lbrace \log_2 \mathrm{det} ( p_{kV}\hat{\mathbf{h}}_{kV}\hat{\mathbf{h}}^H_{kV} + p_{kH}\hat{\mathbf{h}}_{kH}\hat{\mathbf{h}}^H_{kH} +\mathbf{\Upsilon}_{kH}) \right. \nonumber \\
	& \left.- \log_2 \mathrm{det} (\mathbf{\Upsilon}_{kH}) 	\right\rbrace \nonumber\\
	=&\frac{\tau_c -\tau_p}{\tau_c}\mathbb{E}\left\lbrace \log_2 \mathrm{det} \left(  \!\!\mathbf{I}_2 + \mathbf{P}_{k}\hat{\mathbf{H}}_{k} \left( \sum_{l = k+1}^K   {p_{lV}} \hat{\mathbf{h}}_{lV} \hat{\mathbf{h}}^H_{lV} \right.\right.\right. \nonumber \\
	&+ \left.\left.\left. {p_{lH}} \hat{\mathbf{h}}_{lH} \hat{\mathbf{h}}^H_{lH}+ \sum_{l = 1}^K     p_{lV} \mathbf{C}^v_l +p_{lH} \mathbf{C}^h_l + \sigma^2_\mathrm{ul} \mathbf{I}_{M}\right)^{-1} \hat{\mathbf{H}}^H_{k}  \right) \!\!	\right\rbrace,
\end{align}
where $\log_2(1 + \mathbf{x}^H \mathbf{A}^{-1} \mathbf{x}) = \log_2 \mathrm{det}( \mathbf{x}\mathbf{x}^H + \mathbf{A} )- \log_2 \mathrm{det}(  \mathbf{A} ) $ is used in $(a)$. In the step $(b)$, we used $\mathbf{\Upsilon}_{kV} = \mathbf{\Upsilon}_{kH} + p_{kH}\hat{\mathbf{h}}_{kH}\hat{\mathbf{h}}^H_{kH}$. Similarly, the achievable uplink sum SE is
\begin{align}
	&R^\mathrm{ul,SIC} = \sum_{l = 1}^K R^\mathrm{ul,SIC}_l \nonumber \\
	=& \sum_{l = 1}^K \frac{\tau_c -\tau_p}{\tau_c} \mathbb{E} \left\lbrace \log_2 (1 + p_{lV}\hat{\mathbf{h}}^H_{lV} \mathbf{\Upsilon}_{lV}^{-1}\hat{\mathbf{h}}_{lV}) \right.\nonumber \\
	&+ \left.  \log_2(1 + p_{lH}\hat{\mathbf{h}}^H_{lH} \mathbf{\Upsilon}_{lH}^{-1}\hat{\mathbf{h}}_{lH} ) \right\rbrace \nonumber \\
	=&\frac{\tau_c -\tau_p}{\tau_c}\mathbb{E}\left\lbrace \log_2 \mathrm{det} ( p_{1V}\hat{\mathbf{h}}_{1V}\hat{\mathbf{h}}^H_{1V} +  \mathbf{\Upsilon}_{1V})- \log_2 \mathrm{det} (\mathbf{\Upsilon}_{KH}) 	\right\rbrace\nonumber \\
	=&\frac{\tau_c -\tau_p}{\tau_c}\mathbb{E}\left\lbrace \log_2 \mathrm{det} \left( \sum_{l = 1}^K\hat{\mathbf{H}}^H_{l} \mathbf{P}_{l} \hat{\mathbf{H}}_{l}   +  \mathbf{\Upsilon}_{KH}\right) - \log_2 \mathrm{det} (\mathbf{\Upsilon}_{KH}) 	\right\rbrace\nonumber \\
	=& \frac{\tau_c -\tau_p}{\tau_c} \mathbb{E}\left\lbrace \log_2 \mathrm{det} \left(  \mathbf{I}_M + \sum_{l = 1}^K\hat{\mathbf{H}}^H_{l} \mathbf{P}_{l} \hat{\mathbf{H}}_{l}     \left(  \sum_{j = 1}^K     p_{jV} \mathbf{C}^v_j \right. \right. \right. \nonumber \\
	& \left. \left. \left. +p_{jH} \mathbf{C}^h_j + \sigma^2_\mathrm{ul} \mathbf{I}_{M}\right)^{-1}	\right)  \right\rbrace.
\end{align}

\subsection{Proof of Lemma \ref{lemmaDLMR}}

The MR precoding matrix is $\mathbf{W}_k 
= \begin{bmatrix}
	\frac{\sqrt{\rho_{kV} }\hat{\mathbf{h}}_{kV}}{\sqrt{\mathrm{tr}\left(\mathbf{\Gamma}_k^v \right)    }}  & 	\frac{\sqrt{\rho_{kH} }\hat{\mathbf{h}}_{kH}}{\sqrt{\mathrm{tr}\left(\mathbf{\Gamma}_k^h \right)    }} 
\end{bmatrix} = [\mathbf{w}_{kV} \mathbf{w}_{kH}]$. The first expectation to calculate in \eqref{eq:rate-expression} is
\begin{align}
	\mathbb{E}\left\lbrace  \mathbf{H}_k \mathbf{W}_k \right\rbrace &= \mathbb{E}\left\lbrace  \left( \hat{\mathbf{H}}_k + \tilde{\mathbf{H}}_k \right) \mathbf{W}_k \right\rbrace = \mathbb{E}\left\lbrace  \hat{\mathbf{H}}_k \mathbf{W}_k \right\rbrace  + \mathbb{E}\left\lbrace  \tilde{\mathbf{H}}_k \mathbf{W}_k  \right\rbrace \nonumber \\
	&= \mathbb{E}\left\lbrace  \hat{\mathbf{H}}_k {\mathbf{W}}_k \right\rbrace = \mathbb{E}\left\lbrace\begin{bmatrix}
		\hat{\mathbf{h}}^H_{kV}\\
		\hat{\mathbf{h}}^H_{kH}
	\end{bmatrix}\left[ \mathbf{w}_{kV} \mathbf{w}_{kH}\right] \right\rbrace \nonumber \\
&=\mathbb{E}\left\lbrace \begin{bmatrix}
		\hat{\mathbf{h}}^H_{kV} {\mathbf{w}}_{kV} & \hat{\mathbf{h}}^H_{kV} {\mathbf{w}}_{kH} \\
		\hat{\mathbf{h}}^H_{kH} {\mathbf{w}}_{kV} & \hat{\mathbf{h}}^H_{kH} {\mathbf{w}}_{kH}
	\end{bmatrix}\right\rbrace \nonumber \\
& = \begin{bmatrix}
		\sqrt{\rho_{kV}\mathrm{tr}\left( \mathbf{\Gamma}^{v}_{k}\right)} & 0 \\
		0 &\sqrt{\rho_{kH}\mathrm{tr}\left( \mathbf{\Gamma}^{h}_{k}\right)}
	\end{bmatrix}.
\end{align}
Then, the second expectation in \eqref{eq:rate-expression}  is
\begin{align}
	&\mathbb{E}\left\lbrace  \mathbf{H}_k \sum_{l=1}^K \mathbf{W}_l\mathbf{W}^H_l  \mathbf{H}^H_k \right\rbrace \nonumber \\
	&= \mathbb{E}\left\lbrace  \left( \hat{\mathbf{H}}_k + \tilde{\mathbf{H}}_k \right)\sum_{l=1}^K \mathbf{W}_l\mathbf{W}^H_l  \left( \hat{\mathbf{H}}_k + \tilde{\mathbf{H}}_k \right)^H\right\rbrace \nonumber\\
	&= \mathbb{E}\left\lbrace  \left( \hat{\mathbf{H}}_k + \tilde{\mathbf{H}}_k \right) \mathbf{W}_k\mathbf{W}^H_k  \left( \hat{\mathbf{H}}_k + \tilde{\mathbf{H}}_k \right)^H\right\rbrace \nonumber \\
	&+ \sum_{\substack{l=1 \\ l\neq k}}^K \mathbb{E}\left\lbrace   {\mathbf{H}}_k  \mathbf{W}_l\mathbf{W}^H_l  {\mathbf{H}}^H_k \right\rbrace.
\end{align}
First, for $l = k$, we have
\begin{align}
	&\mathbb{E}\left\lbrace   \hat{\mathbf{H}}_k   \mathbf{W}_k\mathbf{W}^H_k   \hat{\mathbf{H}}^H_k \right\rbrace \nonumber \\
	&=   \mathbb{E}\left\lbrace \begin{bmatrix}
		\hat{\mathbf{h}}^H_{kV} {\mathbf{w}}_{kV} & \hat{\mathbf{h}}^H_{kV} {\mathbf{w}}_{kH} \\
		\hat{\mathbf{h}}^H_{kH} {\mathbf{w}}_{kV} & \hat{\mathbf{h}}^H_{kH} {\mathbf{w}}_{kH}
	\end{bmatrix}\begin{bmatrix}
		\hat{\mathbf{h}}^H_{kV} {\mathbf{w}}_{kV} & \hat{\mathbf{h}}^H_{kV} {\mathbf{w}}_{kH} \\
		\hat{\mathbf{h}}^H_{kH} {\mathbf{w}}_{kV} & \hat{\mathbf{h}}^H_{kH} {\mathbf{w}}_{kH}
	\end{bmatrix}^H\right\rbrace \nonumber\\
	&= \begin{bmatrix}
		\mathbb{E}\left\lbrace\left| \hat{\mathbf{h}}^H_{kV} {\mathbf{w}}_{kV}\right|^2 + \left| \hat{\mathbf{h}}^H_{kV} {\mathbf{w}}_{kH}\right|^2\right\rbrace& \!\!\!\!\!\!0\\
		\!\!\!\!\!\!0 & \!\!\!\!\!\!\!\!\!\!\!\!\!\!\!\!\!\!\!\!\!\!\!\!\!\!\!\!\!\!\!\!\!\!\!\!\! \mathbb{E}\left\lbrace\left| \hat{\mathbf{h}}^H_{kH} {\mathbf{w}}_{kH}\right|^2 +\left| \hat{\mathbf{h}}^H_{kH} {\mathbf{w}}_{kV}\right|^2\right\rbrace 
	\end{bmatrix},
\end{align}
where
\begin{align}
	\mathbb{E}\left\lbrace \left| \hat{\mathbf{h}}^H_{kV} \hat{\mathbf{h}}_{kV}\right|^2 \right\rbrace &= \left| \mathrm{tr}\left(\mathbf{\Gamma}_k^v\right)  \right|^2 + \mathrm{tr}\left(\mathbf{\Gamma}_k^v\mathbf{\Gamma}_k^v\right) \nonumber \\
	& = \left| \mathrm{tr}\left( \mathbf{\Gamma}_k^v\right) \right|^2 + \mathrm{tr}\left( \mathbf{\Gamma}_k^v \left( \mathbf{R}_{kV} - \mathbf{C}^{v}_{k}\right) \right) , \\
	\mathbb{E}\left\lbrace \left| \hat{\mathbf{h}}^H_{kV} \hat{\mathbf{h}}_{kH}\right|^2 \right\rbrace &=\mathbb{E}\left\lbrace \left| \hat{\mathbf{h}}^H_{kH} \hat{\mathbf{h}}_{kV}\right|^2 \right\rbrace \nonumber \\
	&=   \mathrm{tr}\left( \mathbf{\Gamma}_k^v\mathbf{\Gamma}_k^h\right) = \mathrm{tr}\left( \mathbf{\Gamma}_k^v \left( \mathbf{R}_{kH} - \mathbf{C}^{h}_{k}\right) \right) \nonumber \\
	&= \mathrm{tr}\left( \mathbf{\Gamma}_k^h \left( \mathbf{R}_{kV} - \mathbf{C}^{v}_{k}\right) \right), \\
	\mathbb{E}\left\lbrace \left| \hat{\mathbf{h}}^H_{kH} \hat{\mathbf{h}}_{kH}\right|^2 \right\rbrace &= \left| \mathrm{tr}\left(\mathbf{\Gamma}_k^h\right)  \right|^2 + \mathrm{tr}\left(\mathbf{\Gamma}_k^h\mathbf{\Gamma}_k^h\right) \nonumber \\
	&= \left| \mathrm{tr}\left( \mathbf{\Gamma}_k^h\right) \right|^2 + \mathrm{tr}\left( \mathbf{\Gamma}_k^h \left( \mathbf{R}_{kH} - \mathbf{C}^{h}_{k}\right) \right).
\end{align}

For $l=k$, the estimation error related part is 
\begin{align}
	&\mathbb{E}\left\lbrace   \tilde{\mathbf{H}}_k   \mathbf{W}_k\mathbf{W}^H_k   \tilde{\mathbf{H}}^H_k \right\rbrace \nonumber \\ 
	&= \begin{bmatrix}
		\mathbb{E}\left\lbrace\left| \tilde{\mathbf{h}}^H_{kV} {\mathbf{w}}_{kV}\right|^2 + \left| \tilde{\mathbf{h}}^H_{kV} {\mathbf{w}}_{kH}\right|^2\right\rbrace& \!\!\! 0\\
	   \!\!\!	0 & \!\!\!\!\!\!\!\!\!\!\!\!\!\!\!\!\!\!\!\!\!\!\!\!\!\!\!\!\!\! \mathbb{E}\left\lbrace\left| \tilde{\mathbf{h}}^H_{kH} {\mathbf{w}}_{kH}\right|^2 +\left| \tilde{\mathbf{h}}^H_{kH} {\mathbf{w}}_{kV}\right|^2\right\rbrace 
	\end{bmatrix}
\end{align}
with $\mathbb{E}\left\lbrace \left| \tilde{\mathbf{h}}^H_{kV} \hat{\mathbf{h}}_{kV}\right|^2 \right\rbrace =  \mathrm{tr}\left(\mathbf{\Gamma}_k^v  \mathbf{C}^{v}_{k}\right)$, $\mathbb{E}\left\lbrace \left| \tilde{\mathbf{h}}^H_{kV} \hat{\mathbf{h}}_{kH}\right|^2 \right\rbrace =   \mathrm{tr}\left( \mathbf{\Gamma}_k^h \mathbf{C}^{v}_{k}\right)$, $\mathbb{E}\left\lbrace \left| \tilde{\mathbf{h}}^H_{kH} \hat{\mathbf{h}}_{kV}\right|^2 \right\rbrace =   \mathrm{tr}\left( \mathbf{\Gamma}_k^v \mathbf{C}^{h}_{k}\right)$ and $\mathbb{E}\left\lbrace \left| \tilde{\mathbf{h}}^H_{kH} \hat{\mathbf{h}}_{kH}\right|^2 \right\rbrace =  \mathrm{tr}\left(\mathbf{\Gamma}_k^h \mathbf{C}^{h}_{k}\right)$. Putting them together gives the result
\begin{align}
	&\!\!\!\!\mathbb{E}\left\lbrace   {\mathbf{H}}_k   \mathbf{W}_k \mathbf{W}^H_k   {\mathbf{H}}^H_k \right\rbrace \nonumber \\
	&\!\!\!\!=    \begin{bmatrix}
		\rho_{kV} \mathrm{tr}\left(\mathbf{\Gamma}_k^v\right)  + \frac{\rho_{kV} \mathrm{tr}\left(\mathbf{\Gamma}_k^v \mathbf{R}_{kV} \right) }{\mathrm{tr}\left(\mathbf{\Gamma}_k^v  \right)} + \frac{\rho_{kH} \mathrm{tr}\left(\mathbf{\Gamma}_k^h \mathbf{R}_{kV} \right) }{\mathrm{tr}\left(\mathbf{\Gamma}_k^h  \right)} \!\!\!\!\!\!\!\!\!\! & \!\!\!\! 0 \\
		\!\!\!\!\!\!\!\!\!\!\!\!\!\!\!\!\!\!\!\!\!\!\!\!\!\!\!\!\!\!\!\!\!\!\!\!\!\!\!\!\!\!\!\!0 \!\!\!\!& \!\!\!\! \!\!\!\!\!\!\!\!\!\!\!\!\!\!\!\!\!\!\!\!\!\!\!\!\!\!\!\!\!\!\!\!\!\!\!\!\!\!\!\!\!\!\!\!\!\!\!\!\!\!\!\!	\rho_{kH}  \mathrm{tr}\left(\mathbf{\Gamma}_k^h\right)  + \frac{\rho_{kH} \mathrm{tr}\left(\mathbf{\Gamma}_k^h \mathbf{R}_{kH} \right) }{\mathrm{tr}\left(\mathbf{\Gamma}_k^h  \right)} + \frac{\rho_{kV} \mathrm{tr}\left(\mathbf{\Gamma}_k^v \mathbf{R}_{kH} \right) }{\mathrm{tr}\left(\mathbf{\Gamma}_k^v  \right)} 
	\end{bmatrix}\!.
\end{align}

For $l \neq k$, we have
\begin{align}
	&\mathbb{E}\left\lbrace   \hat{\mathbf{H}}_k   \mathbf{W}_l\mathbf{W}^H_l   \hat{\mathbf{H}}^H_k \right\rbrace \nonumber \\
	&=  \mathbb{E}\left\lbrace \begin{bmatrix}
		\left| \hat{\mathbf{h}}^H_{kV} {\mathbf{w}}_{lV}\right|^2 +\left| \hat{\mathbf{h}}^H_{kV} {\mathbf{w}}_{lH}\right|^2   & 0\\
		0 & \left| \hat{\mathbf{h}}^H_{kH} {\mathbf{w}}_{lH}\right|^2 +\left| \hat{\mathbf{h}}^H_{kH} {\mathbf{w}}_{lV}\right|^2
	\end{bmatrix}\right\rbrace,
\end{align}
where $\mathbb{E}\left\lbrace \left| \hat{\mathbf{h}}^H_{kV} \hat{\mathbf{h}}_{lV}\right|^2 \right\rbrace =  \mathrm{tr}\left(  \mathbf{\Gamma}_l^v \left( \mathbf{R}_{kV} - \mathbf{C}^{v}_{k}\right) \right)$, 
	$\mathbb{E}\left\lbrace \left| \hat{\mathbf{h}}^H_{kV} \hat{\mathbf{h}}_{lH}\right|^2 \right\rbrace   =   \mathrm{tr}\left( \mathbf{\Gamma}_l^h \left( \mathbf{R}_{kV} - \mathbf{C}^{v}_{k}\right)\right)$, 
	$\mathbb{E}\left\lbrace \left| \hat{\mathbf{h}}^H_{kH} \hat{\mathbf{h}}_{lV}\right|^2 \right\rbrace =   \mathrm{tr}\left(  \mathbf{\Gamma}_l^v \left( \mathbf{R}_{kH} - \mathbf{C}^{h}_{k}\right)\right)$, and 
	$\mathbb{E}\left\lbrace \left| \hat{\mathbf{h}}^H_{kH} \hat{\mathbf{h}}_{lH}\right|^2 \right\rbrace =  \mathrm{tr}\left(  \mathbf{\Gamma}_l^h \left( \mathbf{R}_{kH} - \mathbf{C}^{h}_{k}\right)\right)$. Also, for $l \neq k$ the estimation error related part is
\begin{align}
	&\mathbb{E}\left\lbrace   \tilde{\mathbf{H}}_k   \mathbf{W}_l\mathbf{W}^H_l   \tilde{\mathbf{H}}^H_k \right\rbrace \nonumber \\
	& =  \mathbb{E}\left\lbrace \begin{bmatrix}
		\left| \tilde{\mathbf{h}}^H_{kV} {\mathbf{w}}_{lV}\right|^2 +\left| \tilde{\mathbf{h}}^H_{kV} {\mathbf{w}}_{lH}\right|^2  & 0\\
		& \left| \tilde{\mathbf{h}}^H_{kH} {\mathbf{w}}_{lH}\right|^2 +\left| \tilde{\mathbf{h}}^H_{kH} {\mathbf{w}}_{lV}\right|^2 
	\end{bmatrix}\right\rbrace
\end{align}
with $\mathbb{E}\left\lbrace \left| \tilde{\mathbf{h}}^H_{kV} \hat{\mathbf{h}}_{lV}\right|^2 \right\rbrace =  \mathrm{tr}\left(\mathbf{\Gamma}_l^v \mathbf{C}^{v}_{k}\right) $,
	$\mathbb{E}\left\lbrace \left| \tilde{\mathbf{h}}^H_{kV} \hat{\mathbf{h}}_{lH}\right|^2 \right\rbrace =   \mathrm{tr}\left(\mathbf{\Gamma}_l^h \mathbf{C}^{v}_{k}\right), $
	$\mathbb{E}\left\lbrace \left| \tilde{\mathbf{h}}^H_{kH} \hat{\mathbf{h}}_{lV}\right|^2 \right\rbrace =   \mathrm{tr}\left(\mathbf{\Gamma}_l^v \mathbf{C}^{h}_{k}\right), $
	$\mathbb{E}\left\lbrace \left| \tilde{\mathbf{h}}^H_{kH} \hat{\mathbf{h}}_{lH}\right|^2 \right\rbrace =  \mathrm{tr}\left(\mathbf{\Gamma}_l^h \mathbf{C}^{h}_{k}\right)$ . Arranging the terms  for the case $l \neq k$ gives
\begin{align}
	&\mathbb{E}\left\lbrace   {\mathbf{H}}_k   \mathbf{W}_l\mathbf{W}^H_l   {\mathbf{H}}^H_k \right\rbrace \nonumber \\
	&=    \begin{bmatrix}
		\frac{\rho_{lV} \mathrm{tr}\left(\mathbf{\Gamma}_l^v \mathbf{R}_{kV} \right) }{\mathrm{tr}\left(\mathbf{\Gamma}_l^v  \right)} + \frac{\rho_{lH} \mathrm{tr}\left(\mathbf{\Gamma}_l^h \mathbf{R}_{kV} \right) }{\mathrm{tr}\left(\mathbf{\Gamma}_l^h  \right)} & 0 \\
		0& \frac{\rho_{lH} \mathrm{tr}\left(\mathbf{\Gamma}_l^h \mathbf{R}_{kH} \right) }{\mathrm{tr}\left(\mathbf{\Gamma}_l^h  \right)} + \frac{\rho_{lV} \mathrm{tr}\left(\mathbf{\Gamma}_l^v \mathbf{R}_{kH} \right) }{\mathrm{tr}\left(\mathbf{\Gamma}_l^v  \right)} 
	\end{bmatrix}.
\end{align}
Substituting these terms into \eqref{eq:rate-expression} gives the result in Lemma \ref{lemmaDLMR}.

\bibliographystyle{IEEEtran}
\bibliography{IEEEabrv,refs}

\begin{IEEEbiography}[{\includegraphics[width=1in,height=1.25in,clip,keepaspectratio]{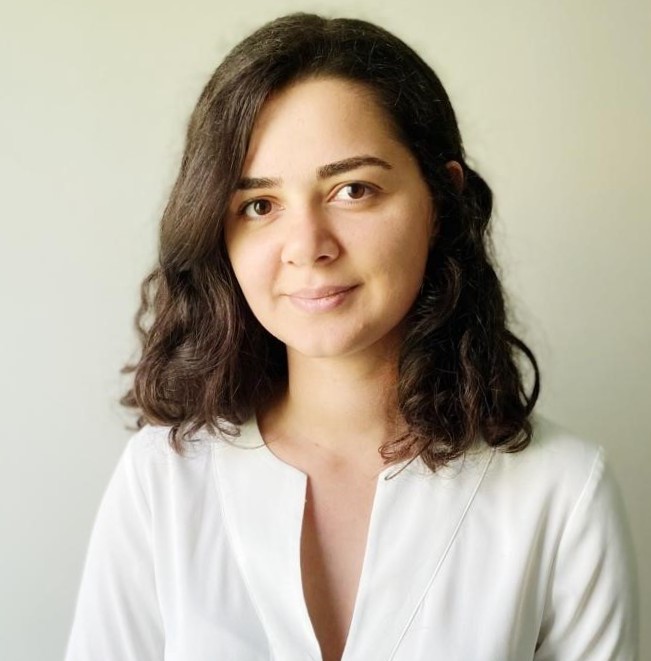}}]{Özgecan Özdogan} received the B.Sc. and
	M.Sc. degrees in electronics and communication
	engineering from the Izmir Institute of Technology,
	Turkey, in 2015 and 2017, respectively. She received the Ph.D. degree in communication systems from Linköping University
	(LiU), Sweden, in 2022. She is now with Ericsson Research, Linköping, Sweden.
\end{IEEEbiography}

\begin{IEEEbiography}[{\includegraphics[width=1in,height=1.25in,clip,keepaspectratio]{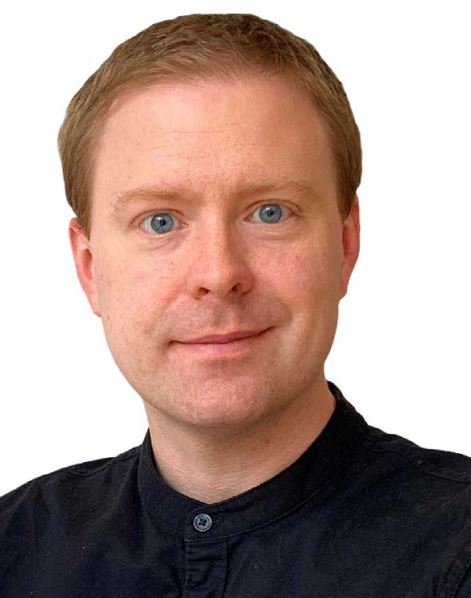}}]{Emil Björnson}
	(Fellow, IEEE) received the M.S.
	degree in engineering mathematics from Lund University, Sweden, in 2007, and the Ph.D. degree in
	telecommunications from the KTH Royal Institute
	of Technology, Sweden, in 2011.
	From 2012 to 2014, he was a Post-Doctoral
	Researcher at the Alcatel-Lucent Chair on Flexible
	Radio, SUPELEC, France. From 2014 to 2021,
	he held different professor positions at Linköping
	University, Sweden. From 2020 to 2021, he was a
	part-time Visiting Full Professor at the KTH. Since
	2022, he has been a Tenured Full Professor of wireless communication at
	the KTH. He has authored the textbooks Optimal Resource Allocation in
	Coordinated Multi-Cell Systems (2013), Massive MIMO Networks: Spectral,
	Energy, and Hardware Efficiency (2017), and Foundations of User-Centric
	Cell-Free Massive MIMO (2021). He is dedicated to reproducible research
	and has made a large amount of simulation code publicly available. He has
	performed MIMO research for over 15 years, his articles have received more
	than 17000 citations, and he has filed more than 20 patent applications. He is
	a Co-Host of the Podcast Wireless Future and has a popular YouTube channel.
	He performs research on MIMO communications, radio resource allocation,
	machine learning for communications, and energy efficiency.
	Dr. Björnson has been a member of the Online Editorial Team of the
	IEEE TRANSACTIONS ON WIRELESS COMMUNICATIONS since 2020. He is
	a Digital Futures Fellow and a Wallenberg Academy Fellow. He has
	received the 2014 Outstanding Young Researcher Award from IEEE ComSoc
	EMEA, the 2015 Ingvar Carlsson Award, the 2016 Best Ph.D. Award from
	EURASIP, the 2018 IEEE Marconi Prize Paper Award in Wireless Communications, the 2019 EURASIP Early Career Award, the 2019 IEEE Communications Society Fred W. Ellersick Prize, the 2019 IEEE Signal Processing
	Magazine Best Column Award, the 2020 Pierre-Simon Laplace Early Career
	Technical Achievement Award, the 2020 CTTC Early Achievement Award,
	and the 2021 IEEE ComSoc RCC Early Achievement Award. He also
	coauthored articles that received Best Paper Awards at the conferences,
	including WCSP 2009, the IEEE CAMSAP 2011, the IEEE SAM 2014, the
	IEEE WCNC 2014, the IEEE ICC 2015, and WCSP 2017. He has been
	on the Editorial Board of the IEEE TRANSACTIONS ON COMMUNICATIONS
	since 2017. He has been an Area Editor of IEEE Signal Processing Magazine
	since 2021.
\end{IEEEbiography}

\end{document}